\documentclass[prb,amsmath,amssymb,twocolumn]{revtex4-2}
\usepackage{graphicx}
\usepackage{tcolorbox}
\usepackage{epsfig}
\usepackage{bbold}
\usepackage{bm}
\usepackage{rotating}
\usepackage{multirow}
\usepackage{xspace}
\usepackage[english]{babel}
\usepackage{hyperref}
\usepackage{ulem}

\def\a{{\alpha}}

\def\w{\omega}
\def\bk{{\bf k}}
\def\bkq{{{\bf k}+ {\bf q}}}

\def\br{{\bf r}}
\def\bq{{\bf q}}

\newcommand{\ket}[1]{ | #1 \rangle }
\newcommand{\bra}[1]{ \langle #1 | }

\def\>{\rangle}
\def\<{\langle}
\def\D{\partial}

\def\k{\kappa}

\begin{document}

\title{Nonlinear electron-phonon coupling drives light-induced symmetry switching in charge-density waves}

\author{Christoph Emeis}
\author{Fabio Caruso}
\affiliation{Institut f\"ur Theoretische Physik und Astrophysik, Christian-Albrechts-Universit\"at zu Kiel, Kiel, Germany} 

\begin{abstract}
Ultrafast optical excitation in charge-density wave (CDW) crystals can
transiently suppress long-range order, driving the lattice toward higher
symmetry on femtosecond timescales.  Here, we formulate and implement a
first-principles theory of light-induced melting of CDW order. The approach is
based on the structural dynamics in the Heisenberg picture, and it explicitly
accounts for quartic lattice anharmonicities, nonlinear electron-phonon
interactions, and photoexcitation-induced modifications of the potential energy
surface.  We illustrate these concepts through first-principles calculations of
the ultrafast melting of CDW order in monolayer TiSe$_2$ -- a prototypical CDW
crystal with a 2$\times$2 structural reconstruction. The simulations are in
good agreement with existing experiments, and they capture the defining
features of CDW melting, such as the damped coherent structural motion, the
transient renormalization of the soft mode, and the restoration of CDW order
over timescales of a few picoseconds.  Besides identifying nonlinear electron-phonon interactions as the primary mechanism driving symmetry switching in CDW systems, our work establishes a generally applicable theoretical framework to treat quartic anharmonicities and light-induced phase transitions in first-principles ultrafast dynamics simulations.
\end{abstract}

\maketitle

Photoexcitation can strongly modify the potential energy surface (PES) of solids,
providing the trigger for coherent structural
motion \cite{RevModPhys.93.041002}. Under intense driving fields, this mechanism
can further induce light-induced phase transitions involving transient
modifications of crystal symmetry \cite{sie2019ultrafast}. Ultrafast
control of crystal symmetry may serve as a handle to influence a variety of
materials properties on femtosecond timescales, including topology
\cite{bao2022light}, optical \cite{PhysRevLett.87.237401} and transport
coefficients \cite{doi:10.1126/science.1241591}, ferroelectricity
\cite{PhysRevLett.118.197601}, and magnetism \cite{PhysRevLett.108.087201}.
Systematically exploiting these principles could lay the physical foundation
for transformative applications in sensing, information storage, or all-optical
electronics \cite{PhysRevLett.98.153905,LiRev2024}. 

Different classes of phase transitions have been observed that are directly
driven by light-induced structural motion, including paraelectric-ferroelectric
switching \cite{doi:10.1126/science.aaw4911,doi:10.1126/science.aaw4913},
paramagnetic-ferromagnetic ordering \cite{doi:10.1126/sciadv.abq2021}, Weyl
semimetal-to-topologically trivial transitions \cite{sie2019ultrafast},
metal-insulator transitions
\cite{PhysRevB.70.161102,rini2007control,PhysRevB.83.195120}, and the melting
of charge-density-wave (CDW) order into a normal metallic state
\cite{doi:10.1126/science.1160778,Tomeljak2009,rohwer2011collapse,PhysRevLett.107.036403,PhysRevLett.107.177402}.
In this work, we focus on CDW melting as a prototypical light-induced phase
transition, although the approach introduced here is 
transferable to other types of driven phase transitions.  In their
low-temperature phase, CDWs display long-range order, which results from the
structural distortion of a reference high-symmetry structure along a
symmetry-breaking normal mode \cite{Rossnagel_2011}. 
At thermal equilibrium, long-range order disappears 
above a critical temperature, the  high-symmetry
structure is restored, and the phase transition is accompanied by the vanishing of
the phonon frequency  \cite{gruner2018density}.  The transient recovery of the high-symmetry
structure can also be realized following a non-thermal pathway, where sample
heating is replaced via ultrafast photoexcitation \cite{Tomeljak2009}.

\begin{figure*}[t]
 \includegraphics[width=0.98\textwidth]{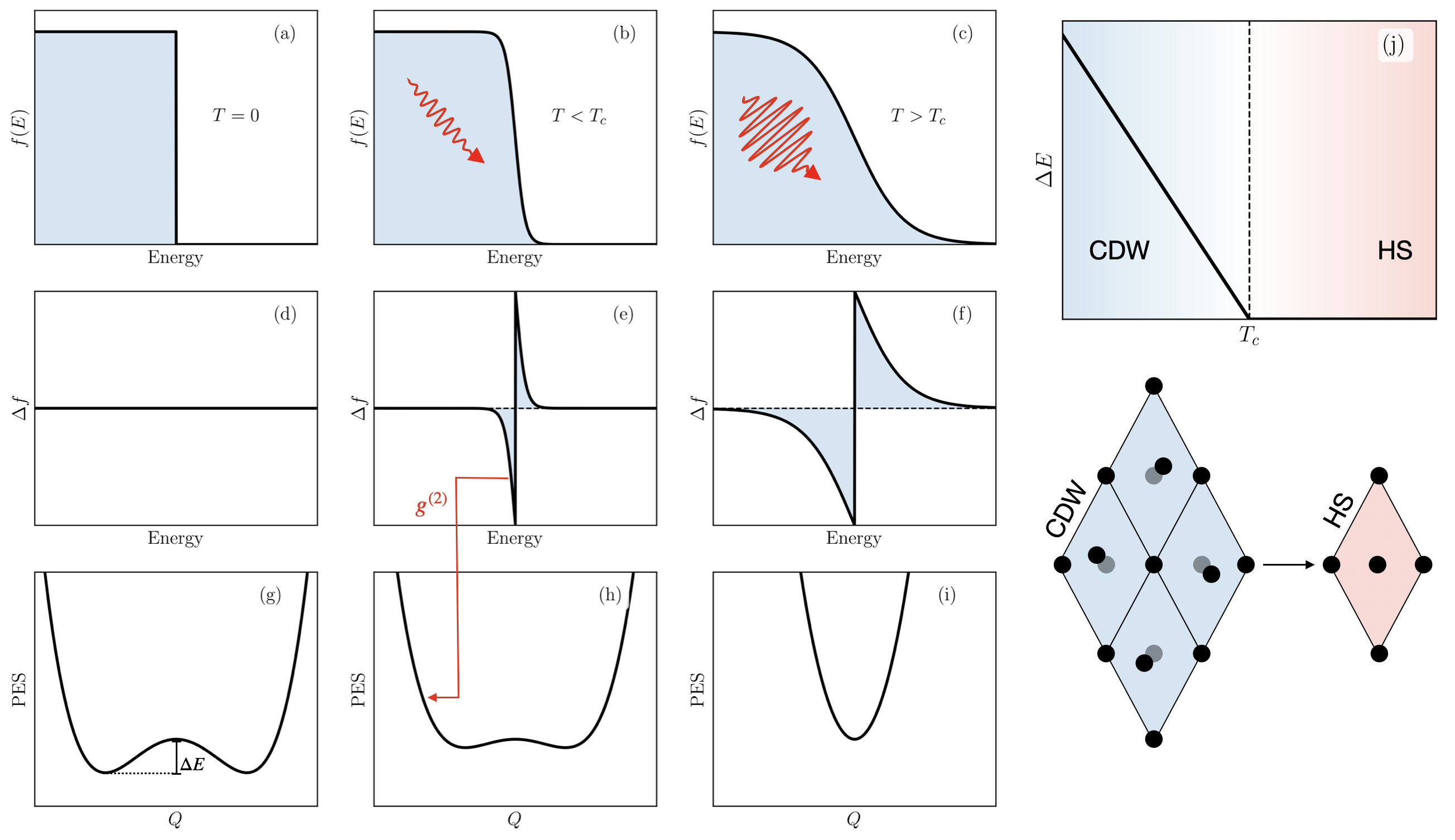}
\caption{ Schematic illustration of key mechanisms responsible for the melting
of CDW order by light absorption.  (a-c)  The electronic distribution function
is modified  by the absorption of a light-pulse.  The changes of the electron
distribution relative to the ground state (d-f)  couple to the lattice via the
nonlinear electron-phonon interactions ($g^{(2)}$), thereby inducing a modification of
the potential energy surface (g-i).  (j) The energy barrier $\Delta E$
separating inequivalent CDW phases is progressively lowered, and vanishes at
the critical excitation density, leading to a transition from CDW to the normal
phase.  } \label{fig:melting}
 \end{figure*}

Figure~\ref{fig:melting} illustrates the key physical principles 
for 
the suppression of CDW order and the transient recovery of a structure 
with higher crystal symmetry. Photoexcitation transfers energy to the electrons, 
resulting in a modification of the electronic occupation function, depicted in
Fig.~\ref{fig:melting}~(a-c) as an increase of the effective electronic temperature $T_e$. 
The ensuing changes of the electron distribution function 
$\Delta f_{n\bk}  = f_{n\bk}(T_e)  - f_{n\bk} (0)$ relative to equilibrium, 
shown in Fig.~\ref{fig:melting}~(d-f), couple to 
the lattice via electron-phonon interactions ($g^{(2)}$), 
leading to a renormalization of the potential energy surface, as illustrated in 
Fig.~\ref{fig:melting}~(g-i). 
The energy barrier $\Delta E$ separating crystal structures with opposite signs of the 
order parameter is suppressed as the effective temperature $T_e$ increases, and vanishes 
above the critical threshold $T_c^e$ (Fig.~\ref{fig:melting}~(j)). Beyond this value, the 
high-symmetry structure becomes energetically more favorable than the lower-symmetry 
CDW structure.  

Pump-probe optical
\cite{Tomeljak2009,PhysRevLett.107.036403,PhysRevB.66.041101,PhysRevLett.83.800,PhysRevB.89.075114,mizukoshi2023ultrafast},
photoemission
\cite{doi:10.1126/science.1160778,rohwer2011collapse,PhysRevLett.105.187401,PhysRevLett.107.177402},
and scattering
\cite{eichberger2010snapshots,PhysRevLett.118.247401,vogelgesang2018phase,kogar2020light,doi:10.1126/sciadv.abf2810}
experiments have been widely employed to track coherent phonons and to monitor
ultrafast phase transitions in CDW crystals, providing insight into
modification of the electronic properties and crystal structure.  Earlier
experimental studies using these techniques have contributed to identifying
several unique fingerprints of light-induced structural phase transitions.
These include, for example, (i) the transient softening of the vibrational
frequency associated with the symmetry-breaking mode 
\cite{Tomeljak2009}; (ii) a structural dynamics
characterized by a broad frequency spectrum above the critical fluence for the
phase transitions \cite{PhysRevLett.107.036403}; (iii) a slowing down of the structural dynamics at
supercritical fluences \cite{Tomeljak2009,maklar2021nonequilibrium}. 
 
Phenomenological models of light-induced phase transitions, such as the
time-dependent Ginzburg-Landau model \cite{RevModPhys.49.435}, have been widely
employed to interpret experiments \cite{maklar2021nonequilibrium}, or to
analyze universal characteristics of the dynamics
\cite{PhysRevLett.116.080601,PhysRevB.101.174306}. They cast the problem in the
form of an equation of motion for the order parameter, which can be solved
numerically and has been shown to reproduce several key features of the
dynamics.  The main limitation of phenomenological models is the lack of
material-specific details,  which are often essential to identify the origin of
coherent structural motion, relevant dissipation pathways, and the interplay of
electronic and vibrational degrees of freedom. 

First-principles ultrafast dynamics simulations are in principle suitable to
model light-induced structural motion without resorting to empirical
parameters, circumventing the limitations of phenomenological approaches.
Progress has been made along several complementary directions, including
time-dependent self-consistent harmonic approximation
\cite{Monacelli_2021,PhysRevB.103.104305}, machine learning potentials
\cite{stocco2025electric}, constrained density-functional theory (DFT)
\cite{PhysRevB.104.144103}, quantum kinetic equations
\cite{mocatti2025nonequilibrium}, time-dependent DFT
\cite{Lian2020,Xu2022,Nie2023}, coherent phonon dynamics in Heisenberg picture
\cite{Caruso2023,PhysRevX.15.021039,Pan2025}, and non-equilibrium Green
function approaches \cite{StefanucciPRX,Stefanucci2025}.  These concerted
efforts reflect the fundamental importance of developing transferable and
predictive theoretical and computational frameworks to model light-induced
structural motion and phase transitions \cite{Caruso_2026}.

The application of ab initio many-body techniques to electron-phonon
interactions poses several key challenges when modeling light-induced structural 
phase transitions of the type illustrated in Fig.~\ref{fig:melting}. 
First, the relevant regions of the potential energy
surface are often strongly anharmonic and feature dynamically unstable normal
modes -- that is, modes characterized by imaginary phonon frequencies. In this
regime, the harmonic approximation fails to adequately describe the lattice
dynamics.
Additionally, the formalism of second-quantization cannot be introduced, as it
requires positive-definite phonon frequencies.  Second, photoexcited carriers
transiently renormalize the potential energy surface through electron–phonon
coupling beyond linear order, introducing nuclear forces that evolve in time as
the electronic distribution relaxes.  Treating these effects within a unified,
fully first-principles many-body description of coupled carrier and lattice
dynamics remains largely an open problem. 
 
In this manuscript,  we develop a theoretical framework to describe the
suppression of CDW order induced by intense driving fields.  After introducing
a unitary transformation of the lattice Hamiltonian, we formulate the lattice
dynamics in Heisenberg picture, explicitly retaining quartic anharmonicities
and electron-phonon interactions up to second order.  The key result of our
work is an equation of motion for light-induced structural dynamics and phase
transitions applicable to CDW crystals and other compounds.
 This approach enables the first-principles 
description of light-induced phase transitions, circumventing the limitations 
of the time-dependent Ginzburg-Landau model. Additionally, our study demonstrates 
that light-induced structural dynamics in CDW crystals is governed by nonlinear 
electron-phonon interactions, whereas ordinary linear coupling vanishes by symmetry. 
To illustrate the modelling capability of this 
approach, we conduct first-principles calculations of the melting of CDW order
in monolayer TiSe$_2$ -- a transition metal dichalcogenide with  a $2\times 2$
CDW reconstruction \cite{Chen2015}.  
Besides reproducing several key characteristics of the
structural motion associated with light-induced phase transitions, our work
introduces a general strategy to handle anharmonicities and 
dynamically unstable phonons within an ab-initio many-body framework. 

The manuscript is organized as follows.  In Sec.~\ref{sec:mod}, we illustrate
the key concepts for the theoretical description of the lattice dynamics for a
one-dimensional model. These ideas are extended to first-principles
calculations in Sec.~\ref{sec:1p}. 
In Sec.~\ref{sec:tise2}, we apply these concepts to describe the light-induced
melting of CDW order.  Discussion and conclusions 
 are reported in Sec.~\ref{sec:disc} and \ref{sec:conc}.
 
\begin{figure*}[t]
\includegraphics[width=0.98\textwidth]{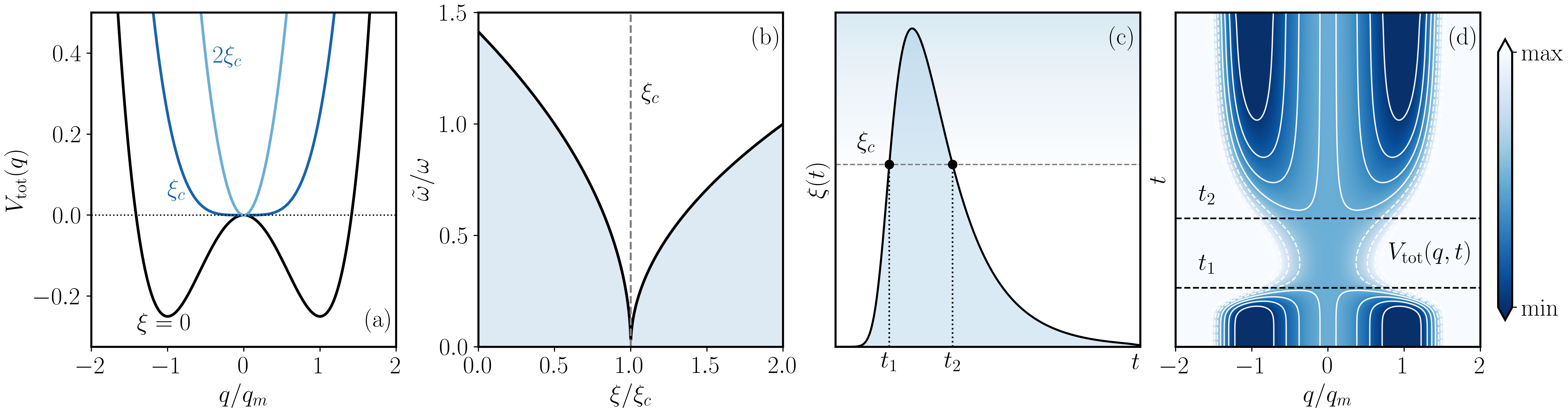}
\caption{(a) Double-well potential $V(q)$ (black) as a function of the position
$q$ for the model Hamiltonian specified by Eq.~\eqref{eq:H}, and its
modification $V(q) + \Delta V$ for interaction strengths $\xi =\xi_c$ and $\xi
= 2\xi_c$.  (b) Dependence of the vibrational frequency on the interaction
strength (Eq.~\eqref{eq:w}).  (c) Time-dependent interaction strengths for
applications to time-dependent problems.  $t_1$ and $t_2$ delimit the time
interval where $\xi (t) > \xi_c $.  (d)  Time-dependent potential (color map)
as a function of displacement and time.  For $t_1 <t < t_2$, the potential is
modified by the perturbation from a double-well to a single well, reflecting
the transition to a higher-symmetry state.} \label{fig:dw}
\end{figure*}

\section{Dynamics in a 1D double-well potential} \label{sec:mod}

We introduce a one-dimensional model to illustrate the key ingredients of the
theoretical description of time-dependent symmetry switching induced by
nonlinear electron-phonon interactions in CDW crystals. Specifically, we
consider a particle moving in a double-well potential and subject to a
quadratic perturbation. The system is described by the Hamiltonian: 
\begin{align}\label{eq:H}
\hat H =  \frac{\hat p^2}{2} - \frac{\w^2}{2} \hat q^2 +  \frac{\beta}{4} \hat q^4 + \Delta \hat V 
\quad.
\end{align}

$\hat q$ is the position operator and $\hat p$ its conjugate momentum.  $\Delta
\hat V  = \xi \hat q^2$ denotes a quadratic perturbation, 
which mimics the effects of nonlinear (second-order) 
electron-phonon interactions arising from a 
time-dependent carrier density.  
The bare potential
$\hat V(q) = -\w^2 \hat q^2 / 2 + \beta \hat q^4/4$ is illustrated in
Fig.~\ref{fig:dw}~(a) in black, whereas the effect of $\Delta \hat V $ for
different values of the coupling parameter $\xi$ are in colours.  In the
context of the Ginzburg-Landau theory of second-order structural phase
transitions, $\hat q$ denotes the collective coordinate for the nuclear
displacement along a symmetry-breaking structural distortion.  The local
maximum ($q=0$), thus,  corresponds to a reference high-symmetry crystal
structure, while the two minima represent lower-symmetry structures generated
by finite displacements ($q\neq 0$) along a symmetry-breaking mode. 

For $\xi=0$, the system is found in one of the two broken-symmetry minima, with
displacement coordinate $q_m =  \pm \w / \sqrt{\beta}$.  For perturbations
below a critical threshold $\xi< \xi_c = \w^2/2$, the total potential $V (q) +
\Delta V$  retains a double-well character and no change of symmetry occurs. 
Conversely, when the critical threshold is
exceeded, $\xi>\xi_c$, the total potential exhibits a unique minimum at $q=0$,
bringing the system to a higher-symmetry structure.  The vibrational frequency
as a function of the interaction strength, illustrated in
Fig.~\ref{fig:dw}~(b),  obeys:
\begin{align}
\label{eq:w}
\tilde\w (\xi) = \sqrt{2} \sqrt{\w^2- 2\xi} \theta (\xi_c-\xi) + \sqrt {2\xi-\w^2} \theta(\xi-\xi_c) \quad. 
\end{align}

It vanishes at $\xi = \xi_c$, following the characteristic trends expected for
second-order phase transitions within the local Landau-Ginzburg theory.
\cite{goldenfeld2018lectures}  

The model is extended to time-dependent perturbations by introducing a
time-dependent interaction strength $\xi(t)$. In the context of pump-probe
spectroscopy, for example, the interaction strength can arise from third-order
anharmonic effects \cite{PhysRevB.89.220301,PhysRevLett.118.054101} or from
the coupling of the lattice to a photoexcited electron density
\cite{Caruso2023}. Here, we consider the latter, as it is the primary mechanism
for light-induced phase transitions in CDW materials.  The interaction strength
$\xi(t)$ follows the characteristic trend illustrated in Fig.~\ref{fig:dw}~(c),
where the increase and decrease of $\xi$ reflect excitation and thermalization of
photocarriers, respectively.  If the interaction strength exceeds the threshold for the phase
transition ($\xi>\xi_c$) for some time interval $ t_1 < t < t_2 $, the
time-dependent potential transiently switches from double-well to single well
(Fig.~\ref{fig:dw}~(d)). The ensuing non-equilibrium dynamics of the system
consists of coherent oscillations around the high-symmetry crystal structure,
and the system is dynamically driven into a high-symmetry state.  This
mechanism provides a minimal picture of the ultrafast melting of CDW order.
However, it neglects inhomogeneity and the emergence of long-range correlations
\cite{RevModPhys.49.435}. 

To study the dynamics arising from the time-dependent Hamiltonian specified by
Eq.~\eqref{eq:H}, with a time-dependent interaction strength $\xi(t)$, it is
convenient to introduce the unitary transformation: 
\begin{align} \label{eq:tr}
\hat q \rightarrow \hat q 
+ \w/\sqrt{\beta}\quad,\quad \hat p \rightarrow \hat p\quad.
\end{align}
This transformation shifts the origin of the coordinate system ($q=0$) to the
right minimum of the quartic potential.  Apart for a constant additive term,
the Hamiltonian becomes: 
\begin{align} 
\hat H=\frac{\hat p^2}{2}
+\frac{\Omega^2 \hat q^2}{2}  &+ \Omega \sqrt{\frac{\beta}{2}} \hat q^3 
\label{eq:H1}
+
\frac{\beta}{4} \hat q^4 
\\ 
&+
 \xi(t) \left [ \hat q^2 + \hat q \Omega\frac{\sqrt{2}}{\sqrt\beta } \right]
\quad,
\nonumber
\end{align}
where $\Omega = \w/\sqrt{2}$.  The first two terms can be identified with the
Hamiltonian of the quantum harmonic oscillators, enabling the introduction of
bosonic creation and annihilation operators, $\hat a^\dagger$ and $\hat a$: 
\begin{align}
\hat q = \left(\frac{\hbar}{2\Omega}\right)^\frac{1}{2} (\hat a + \hat a^\dagger) 
\quad,\quad
\hat p = i \left(\frac{ \hbar\Omega} {2}\right )^\frac{1}{2}  (\hat a - \hat a^\dagger)\quad. 
\end{align}
Substitution into Eq.~\eqref{eq:H1} yields:
\begin{align}  
\hat H = \hbar\Omega \hat a^\dagger \hat a 
&+ \frac{1}{4}\sqrt{\frac{\beta \hbar^3}{\Omega}} \hat Q^3
+\frac{\beta \hbar^2}{16 \Omega^2} \hat Q^4 \label{eq:H2}\\ 
&+ \frac{\hbar \xi(t) }{ 2\Omega}
\left[
\hat Q^2 +
2 \hat Q \left( \frac{\Omega^3}{\hbar\beta} \right)^{1/2}
\right]
\quad. 
\nonumber
\end{align}
with the abbreviation $\hat Q = \hat a+ \hat a^\dagger$.  The advantage of
rewriting the Hamiltonian in this form is that the dynamics of the system can
be investigated  from the direct solution of the Heisenberg equation of motion
for the displacement operator  $\partial_t^2  \hat Q = -\hbar^{-2} [[\hat Q,
\hat H],\hat H]$, where $\partial_t =  \partial / \partial t$
\cite{Caruso2023}. In particular, the nested commutators can be evaluated via
repeated application of the ordinary bosonic commutation relations, leading to
the following equation of motion for the lattice displacement:  
\begin{align} \label{eq:eom}
&\quad\quad\partial_t^2 Q 
+\gamma \partial_t Q + \Omega^2 Q = D(t) \quad,
\end{align} 

where $\gamma$ is the  damping rate \cite{Pan2025,Stefanucci2025}, and we
introduced the function: 
\begin{align}
D(t) = - \frac{3}{2} \sqrt{{\hbar \Omega \beta}} Q^2
- \frac{\hbar \beta}{2\Omega} Q^3
- 2\xi (t)
\left [   Q  +
\left(\frac{{\Omega^3}}{{\hbar\beta}} \right)^{\frac{1}{2}}
\right]
\quad. 
\end{align}

Details of the derivations are given in Appendix~\ref{sec:1Dmodelder}. 
Here, we replaced operators with their expectation values, in the
semiclassical approximation $\langle \hat Q^n \rangle = Q^n$ commonly used in
non-linear phononics models \cite{PhysRevB.89.220301}.
In short, Eq.~\eqref{eq:eom} reformulates the problem of the dynamics of a
particle in a time-dependent double well in the form of a second-order
differential equation that can be solved via ordinary time-stepping algorithms. 
For consistency, we note that in the limit of small displacement, 
Eq.~\eqref{eq:eom} reduces to the displacive excitation of coherent phonons \cite{PhysRevX.15.021039}, 
reflecting a lattice dynamics characterized by small damped oscillations confined
to one of the two broken-symmetry minima of the PES. 
\section{Light-induced perturbation from the electron-phonon interaction} \label{sec:1p}
We here proceed to formulate a theoretical framework suitable for capturing
the coherent lattice motion and symmetry switching occurring in CDW materials
at high driving fluences by explicitly retaining quartic anharmonic terms in
the PES. This is achieved by generalizing the concepts introduced 
in Sec.~\ref{sec:mod}. The procedure involves three steps: in Sec.~\ref{sec:phtr} we introduce a
coordinate transformation of the lattice Hamiltonian to handle
dynamically-unstable normal modes, while retaining the anharmonic character of
the potential; this approach is extended to the first- and second-order
electron-phonon coupling Hamiltonians in Sec.~\ref{sec:ephtr}; finally, the equation of motion for the adiabatic lattice dynamics in CDW materials
in presence of a photoexcited carrier density is derived in Sec.~\ref{sec:heom}
within the Heisenberg picture. The result of this procedure is encoded by
Eqs.~\eqref{eq:EOM1}-\eqref{eq:EOM3}, which recast the structural dynamics upon
photoexcitation in the form a second-order differential equation for the atomic
displacements.

\subsection{Lattice Hamiltonian and dynamical instabilities}\label{sec:phtr}

Following the standard procedure for theoretical description of crystal lattice 
vibrations \cite{10.1093/oso/9780192670083.001.0001}, we consider the expansion of the PES $U$ as a power series in the 
displacements $\Delta \hat \tau_i$ of the atoms from equilibrium:
\begin{align} 
U = U_0 + \sum_{n=2}^{\infty}  \frac{1}{n!} 
\sum_{i_1 \dots i_n}
\left. \frac{\partial^n U} {\partial  
\Delta \tau_{i_1} \dots \partial \Delta \tau_{i_n}}  
\right| _0
\Delta \hat \tau_{i_1} \dots  \Delta \hat \tau_{i_n}\, . 
\label{eq:pes-all}
\end{align}
$i = \{\k,p,\alpha \}$ is a collective index for the $\a$-th 
Cartesian coordinate,  $\k$-th atom, and $p$-th unit cell in 
Born-von-Karman periodic boundary conditions.
The displacements admit a representation in a normal-mode basis: 
\begin{align}
\Delta \tau_i =
\frac{1}{(N_p M_\k) ^{1/2}}
\sum_{\bq\nu} e^{i\bq {\bf R}_p}  e_{\bq\nu}^{\k \a} \hat q_{\bq\nu}
\quad. 
\label{eq:nm}
\end{align}
$\{ e_{\bq\nu}^{\k \a} \}$ denotes the eigenvectors of the dynamical matrix,
$M_\k$ the atomic mass, $N_p$ the number of unit cells in the supercell, and
${\bf R}_p$ a crystal lattice vector.  $\hat q_{\bq\nu}$ is an operator that
quantifies the lattice displacement along the normal-mode $\bq\nu$. 
Substituting Eq.~\eqref{eq:nm} into Eq.~\eqref{eq:pes-all} 
leads to the normal-mode representation of the 
lattice Hamiltonian: 
\begin{align}
\label{eq:nm1}
\hat H &= 
 \frac{1}{2}
\sum_{\bq\nu} 
[
\hat p_{\bq\nu} 
\hat p_{-\bq\nu} 
+ 
\w_{\bq\nu}^2
\hat q_{\bq\nu}
\hat q_{-\bq\nu}
] 
\\ 
& + \frac{1}{3!}
\sum_{\bq_1\dots \bq_3 \atop \nu_1 \dots \nu_3}
\Psi^{(3)}_{\bq_1 \dots \bq_3 \atop \nu_1 \dots \nu_3}
\hat q_{\bq_1\nu_1}
\hat q_{\bq_2\nu_2}
\hat q_{\bq_3\nu_3}
\delta_{\bq_1+ \bq_2+\bq_3}^{\bf G}\nonumber 
\\ 
& + 
\frac{1}{4!}
\sum_{\bq_1\dots \bq_4 \atop \nu_1 \dots \nu_4}
\Psi^{(4)}_{\bq_1 \dots \bq_4 \atop \nu_1 \dots \nu_4} 
\hat q_{\bq_1\nu_1}
\hat q_{\bq_2\nu_2}
\hat q_{\bq_3\nu_3}
\hat q_{\bq_4\nu_4}
\delta_{\bq_1+ \bq_2+\bq_3+\bq_4}^{\bf G} \nonumber \, 
\end{align}
$\hat p_{\bq\nu}$ is the conjugate momentum to $ \hat q_{\bq\nu}$, 
$\w_{\bq\nu}^2$ are the eigenvalues of the dynamical matrix, $\Psi^{(3)}$ and
$\Psi^{(4)}$ are the third- and fourth-order phonon-phonon coupling matrix
elements, respectively.  

Starting from Eq.~\eqref{eq:nm1}, we restrict ourselves to consider crystal
structures in which normal modes can be classified into: (i) a subset of
dynamically stable modes, labeled with $\mathcal S$, for which the eigenvalues
of the dynamical matrix fulfil $\w_{\bq\nu}^2 \ge 0$; (ii)  a subset of
dynamically unstable modes, labeled with $\mathcal U$, which fulfil
$\w_{\bq\nu}^2 < 0$.  For a given mode with indices $\bq\nu$ in the subset
$\mathcal U$, we further assume that the potential energy surface can be
approximately described by a symmetric double-well potential in the form: 
\begin{align}  \label{eq:tmpx}
V_{\bq\nu} = - \frac{| \w_{\bq\nu}|^2 }{2}  
\hat q_{\bq\nu} ^2 + \frac{\beta_{\bq\nu}}{4} 
\hat q_{\bq\nu}^4
\quad.
\end{align}
For simplicity, we considered real-valued normal coordinates $ q_{\bq\nu} =
q_{-\bq\nu}$, which applies to crystal momenta $\bq$ at the edge of the
Brillouin zone \cite{Zacharias2020}.
The approximate PES considered above encodes the degrees of freedom necessary
to capture the emergence of $2\times2\times1$ and $2\times2\times2$ CDW
reconstructions, as observed, for example, in 1T-TiSe$_2$ \cite{DiSalvo1976,Watson2019},
monolayer transition-metal dichalcogenides \cite{Chen2015,Fang2017}, and kagome
metals \cite{Christensen2021,Xiao2023,Yao2024}.
These considerations can
readily be extended to other types of CDW reconstructions by retaining
complex-valued normal coordinates in Eq.~\eqref{eq:tmpx}.  Additionally, while
our work specifically targets soft modes, structural instability, and phase
transitions in CDW compounds, other materials families may further be
described under similar assumptions, as, e.g.,  ferroelectrics or polymorphous
perovskites \cite{zacharias2023anharmonic}.  The phonon Hamiltonian can thus be
partitioned into contribution arising from the subset $\mathcal S$ and
$\mathcal U$ of stable and unstable normal modes:
\begin{align} 
\hat H_{\rm ph} 
&= 
\label{eq:hph0}
\hat H_{\rm ph}^{\mathcal S} 
+ \hat H_{\rm ph}^{\mathcal U} 
\quad, 
\\ 
\hat H_{\rm ph}^{\mathcal S} &= 
\frac{1}{2}\sum_{\bq\nu}^{\mathcal S }  
[
| \hat p_{\bq\nu} |^2
+
\w_{\bq\nu}^2
| \hat q_{\bq\nu}|^2
]
\quad,
\label{eq:HS}
 \\
\hat H_{\rm ph}^{\mathcal U}
&
= \frac{1}{2}\sum_{\bq\nu}^{\mathcal U} 
\left[
 \hat p_{\bq\nu} ^2
-
| \w_{\bq\nu}| ^2
 \hat q_{\bq\nu}^2 
+\frac{\beta_{\bq\nu}}{2} 
 \hat q_{\bq\nu}^4
\right]
\quad. 
\end{align}
For all stable modes in the subset $\mathcal S$, bosonic operators can be 
rewritten in  second-quantization via: 
\begin{align}  
\label{eq:2qQ}
\hat q_{\bq\nu} = \left( \frac{\hbar}{2\w_{\bq\nu}} \right)^{1/2}
(\hat a_{\bq\nu} + \hat a_{-\bq\nu}^\dagger) 
\quad. 
\end{align}
Substitution in Eq.~\eqref{eq:HS} yields the 
usual second-quantized form of the free-phonon Hamiltonian: 
\begin{align} 
\hat H_{\rm ph}^{\mathcal S} 
&= 
\sum_{\bq\nu}^{\mathcal S} 
\hbar \w_{\bq\nu} 
\hat a_{\bq\nu }^\dagger 
\hat a_{\bq\nu }  
\quad. 
\end{align}
This procedure, however, cannot be straightforwardly applied to the unstable
modes, since real-valued phonon frequencies are required.  This issue can be
circumvented by noting that the potential $\hat V_{\bq\nu}$, defined in
Eq.~\eqref{eq:tmpx}, has minima at $q_{\rm min} = \pm | \w_{\bq\nu}| /
\sqrt{\beta_{\bq\nu}}$.  These points, which corresponds to the ground state of
the system at equilibrium, have positive-definite vibrational frequency
$\Omega_{\bq\nu}^2= \left. \frac{\partial^2 V_{\bq\nu}}{ \partial q_{\bq\nu}^2
}\right|_{q=q_{\rm min}}  =
- \frac{\w_{\bq\nu}^2 } {2}$ (with $\w_{\bq\nu}^2<0$).  This suggests to introduce the coordinate transformation: 

\begin{align} \label{eq:coordtr}
\hat q_{\bq\nu} 
\rightarrow \hat {\tilde q}_{\bq\nu} =  \hat q_{\bq\nu}  
- \frac{| \w_{\bq\nu}|}{\sqrt{ \beta_{\bq\nu}}}\quad, 
\end{align}

and to introduce the second-quantized representation for the shifted 
operator: 

\begin{align}
\hat {\tilde q}_{\bq\nu} = 
\left( \frac{\hbar } {2\Omega_{\bq\nu}} \right)^{1/2} (\hat a_{\bq\nu} + \hat a_{-\bq\nu}^\dagger)
\quad.
\end{align}

After few algebraic steps, the non-interacting phonon Hamiltonian 
for the subset of unstable modes can be rewritten as: 

\begin{align}
\hat H_{\rm ph}^{\mathcal U} 
&= 
\sum_{\bq\nu}^{\mathcal U} 
 \hbar \Omega_{\bq\nu}  
\hat a_{\bq\nu }^\dagger
\hat a_{\bq\nu } 
+ \hat V^{(a)} 
\quad,
\end{align} 

where we introduced the anharmonic potential $\hat V^{(a)}$:  
\begin{align}
 \hat V^{(a)}
= 
\sum_{\bq\nu}^{\mathcal U}
\left[ \frac{1}{4} \sqrt{\frac{\beta_{\bq\nu} \hbar^3}{\Omega_{\bq\nu}}} 
\hat Q_{\bq\nu}^3
+ \frac{\beta_{\bq\nu} \hbar ^2 }{16\Omega_{\bq\nu}^2} 
\hat Q_{\bq\nu}^4 \right]
\quad.
\end{align}
In summary, via the coordinate transformation introduced in
Eq.~\eqref{eq:coordtr}, the Hamiltonian is rewritten in the form of quantum
harmonic oscillator with positive-definite vibrational frequencies
$\Omega_{\bq\nu}$. At the same time, the double-well character of the potential
energy surface is preserved at the cost of retaining the third- and
fourth-order anharmonic terms in the potential $\hat V^{(a)}$. 

\subsection{Electron-Phonon interaction}\label{sec:ephtr} 

Following the procedure introduced in Sec.~\ref{sec:phtr}, we handle separately dynamical stable ($\mathcal{S}$) and unstable ($\mathcal{U}$) modes in the definition of the electron-phonon coupling Hamiltonian. 
In Appendix~\ref{sec:AHep}, we show that the  coordinate transformation 
defined in Eq.~\eqref{eq:coordtr} can be
applied to the first- and second-order electron–phonon interaction 
Hamiltonians to properly account for instabilities. This procedure leads to a decomposition of the EPI Hamiltonian into:
\begin{align} \label{eq:HEPItot}
\hat H_{\rm eph}^{(1)} 
&= 
\hat H_{\rm eph}^{(1)\,\mathcal{S}} + 
\hat H_{\rm eph}^{(1)\,\mathcal{U}} 
\\ 
\hat H_{\rm eph}^{(2)}
 \label{eq:HEPItot2}
&=
\hat H_{\rm eph}^{(2)\,\mathcal{SS}} + 
\hat H_{\rm eph}^{(2)\,\mathcal{US}} +
\hat H_{\rm eph}^{(2)\,\mathcal{UU}}
\quad. 
\end{align}
The first and second-order contribution arising solely from unstable 
modes are defined by: 
\begin{widetext}
\begin{align}
\hat H^{(1)\,\mathcal {U}}_{\rm eph} &=
\frac{1}{\sqrt{N_p} }
\sum_{\bq\nu}^{\mathcal U}
\sum_{nm\bk} \tilde g _{nm}^\nu(\bk,\bq)
[ \hat c_{m\bkq}^\dagger \hat c_{n\bk} - f^0_{n\bk} \delta_{nm} \delta_{\bq,0} ]
\left[ \hat Q_{\bq\nu}   +
\left(\frac{ \Omega_{\bq\nu}^3} {\hbar\beta_{\bq\nu}}\right)^{1/2}
\right] \quad, 
\\
\label{eq:HUU}
\hat H^{(2)\,\mathcal {UU}}_{\rm eph} &=
\frac{1}{N_p } \sum_{\bq\nu}^\mathcal{U}
\sum_{\bq'\nu'}^\mathcal{U} \sum_{nm\bk}
\tilde g _{nm}^{\nu\nu'}(\bk,\bq,\bq')
 [ \hat c_{m\bkq+\bq'}^\dagger \hat c_{n\bk} - f^0_{n\bk} \delta_{nm} \delta_{\bq+\bq',0} ]
\\
&\quad\quad\quad
\quad\quad\quad
\quad\quad\quad
\times \left[\hat Q_{\bq\nu} \hat Q_{\bq' \nu'}
+  2
\hat Q_{\bq' \nu'}
\left(\frac{ \Omega^3_{\bq\nu}} {\hbar \beta_{\bq\nu}}\right)^{1/2}
 +
\left(\frac{ \Omega^3_{\bq\nu} \Omega^3_{\bq'\nu'} } {\hbar^2
\beta_{\bq\nu} \beta_{\bq'\nu'}} \right)^{1/2}
 \right]
\quad, 
\nonumber
\end{align}
with matrix elements defined according to:
\begin{align}
\tilde g _{nm}^{\nu}(\bk,\bq) &=
\left(\frac{\hbar} {2\Omega_{\bq\nu}}  \right)^{1/2}
\bra{\psi_{m\bkq+}}
\partial_{\bq\nu}
v_{\rm KS} \ket{\psi_{n\bk}}
\quad,
\\
\tilde g _{nm}^{\nu\nu'}(\bk,\bq,\bq') &=
\left(\frac{\hbar^2} {4 \Omega_{\bq\nu}\Omega_{\bq'\nu'}}  \right)^{1/2}
\bra{\psi_{m\bkq+\bq'}}
\partial_{\bq'\nu'}
\partial_{\bq\nu}
v_{\rm KS} \ket{\psi_{n\bk}}
\quad. \label{eq:tmpx1324}
\end{align}
\end{widetext}
Remaining terms are reported in Appendix~\ref{sec:AHep}.
Equations~\eqref{eq:HEPItot}-\eqref{eq:tmpx1324} extend the definition of the
EPI Hamiltonian to crystal structures characterized by dynamically unstable
modes well-represented by  a quartic double-well potential in the form of
Eq.~\eqref{eq:tmpx}. All matrix elements entering these expressions are
evaluated for the high-symmetry structure.
\subsection{Lattice dynamics in Heisenberg picture}\label{sec:heom}
We determine the structural dynamics induced by optical excitation through
solution of the Heisenberg equation of motion for the position operator:
\begin{align} \label{eq:HEOM}
\D_t^2\hat Q_{\bq\nu} = -\hbar^{-2}
\big[[\hat Q_{\bq\nu}, \hat H], \hat H\big]\quad,
\end{align}
$\hat H$ is the lattice Hamiltonian in presence of EPI up to second order:
\begin{align}
\hat H = 
\hat H_{\rm ph} 
+ \hat H_{\rm eph}^{(1)}
+ \hat H_{\rm eph}^{(2)}
\quad.
\end{align}
We focus on crystal lattices with  
dynamical instabilities for 
which the phonon and electron-phonon Hamiltonians can be 
decomposed as in Eqs.~\eqref{eq:hph0},
\eqref{eq:HEPItot}, and \eqref{eq:HEPItot2}. 
The adiabatic equation of motion for the displacement $Q_{\bq\nu}$ can 
thus be determined via the procedure outlined in Ref.~\cite{Caruso2023}. 
We outline the key steps: 
(i) explicit evaluation of the nested commutators in 
Eq.~\eqref{eq:HEOM} yields an equation of motion linking 
bosonic and fermionic operators; 
(ii) the lattice response to the electrons is treated adiabatically by 
replacing fermionic operators with their expectation value on the state 
$\psi_{\rm el}(t)$, which describe the electrons at time $t$. 
This procedure -- illustrated in detail in Appendix~\ref{sec:AEOM} -- 
leads to the following equation of motion for the displacement: 
\begin{align}
\label{eq:EOM1}
& \D_t^2Q_{\bq\nu} +\gamma_{\bq\nu} \D_t Q_{\bq\nu} +  \Omega_{\bq\nu} ^2 Q_{\bq\nu}
= D_{\bq\nu}(t) \\
&D_{\bq\nu}(t) =
-
\frac{3}{2} ({\beta_{\bq\nu} \hbar\Omega_{\bq\nu}})^{\frac{1}{2}}
Q_{\bq\nu}^2
- \frac{\beta_{\bq\nu}\hbar}{2 \Omega_{\bq\nu}} \hat Q_{\bq\nu}^3
\nonumber
\\ 
&
\quad \quad \quad  \quad 
\quad-
2\xi_{\bq\nu}(t)
\left [
 \hat Q_{\bq\nu}
+  \left(\frac{\Omega_{\bq\nu}^3}{{\hbar\beta_{\bq\nu}}} \right)^{1/2}
\right]
\\
&\xi_{\bq\nu}(t) =
2 N_p^{-1} \frac{\Omega_{\bq\nu}}{\hbar} \sum_{n\bk}
\tilde g_{nn}^{\nu\nu} (\bk,\bq,-\bq)
\Delta f_{n\bk}(t)
\label{eq:EOM3}
\end{align}
Here, $\Delta f_{n\bk}(t)$ denotes the time-dependent change of 
electronic occupation arising, e.g., from photoexcitation. 
Equations~\eqref{eq:EOM1}-\eqref{eq:EOM3} are the central result of this work.
They reformulate the structural dynamics in the direction of the unstable
normal model $\bq\nu$ in the form of a second-order differential equation,
which can be solved via ordinary time-stepping algorithm.  With the exception
of the parameter $\beta_{\bq\nu}$ -- which contains information about the
quartic terms in the potential energy surface --, all quantities can be
obtained from first-principles calculations for the high-symmetry structure in
the unit cell. 
Equations~\eqref{eq:EOM1}-\eqref{eq:EOM3} neglect electronic coherence
($\bra{\psi_{\rm el}(t)} \hat c_{m\bkq}^\dagger  \hat c_{n\bk} \ket{\psi_{\rm
el}(t)} \simeq f_{n\bk}\delta_{nm} \delta_{\bq,0}$).  A detailed symmetry
analysis of the EPI matrix elements, reported in Appendix \ref{sec:sym},
reveals that the first-order electron-phonon coupling does not contribute to
the structural dynamics for symmetry-breaking normal modes -- namely, with
symmetries different from $A_g$ or $A_{1g}$ in the symmetry group of the 
parent high-symmetry structure.  From this point, we deduce that
the lowest-order contribution to light-induced structural motion in CDW
crystals occurs at second-order in the EPI, and are therefore a manifestation
of nonlinear interactions in driven crystals. 

The quantity $\xi_{\bq\nu}$, defined in Eq.~\eqref{eq:EOM3}, plays the role of
the time-dependent coupling strength, defined in analogy to the 1D model of
Sec.~\ref{sec:mod}.  It depends on the second-order EPI matrix element $\tilde
g_{nn}^{\nu\nu} (\bk,\bq,-\bq)$ for the high-symmetry structure and on the
change of electronic occupation $\Delta f_{n\bk}(t)$.  In Appendix \ref{sec:2to1} 
we show that it can alternatively be expressed in terms of the first-order EPI matrix
elements for the CDW phase (see Eq.~\ref{eq:xi_SC}). This form is particularly 
advantageous for first-principles calculations, as it circumvents the 
explicit calculation of the second-order matrix element. 

\begin{figure*}[t]
\includegraphics[width=0.98\textwidth]{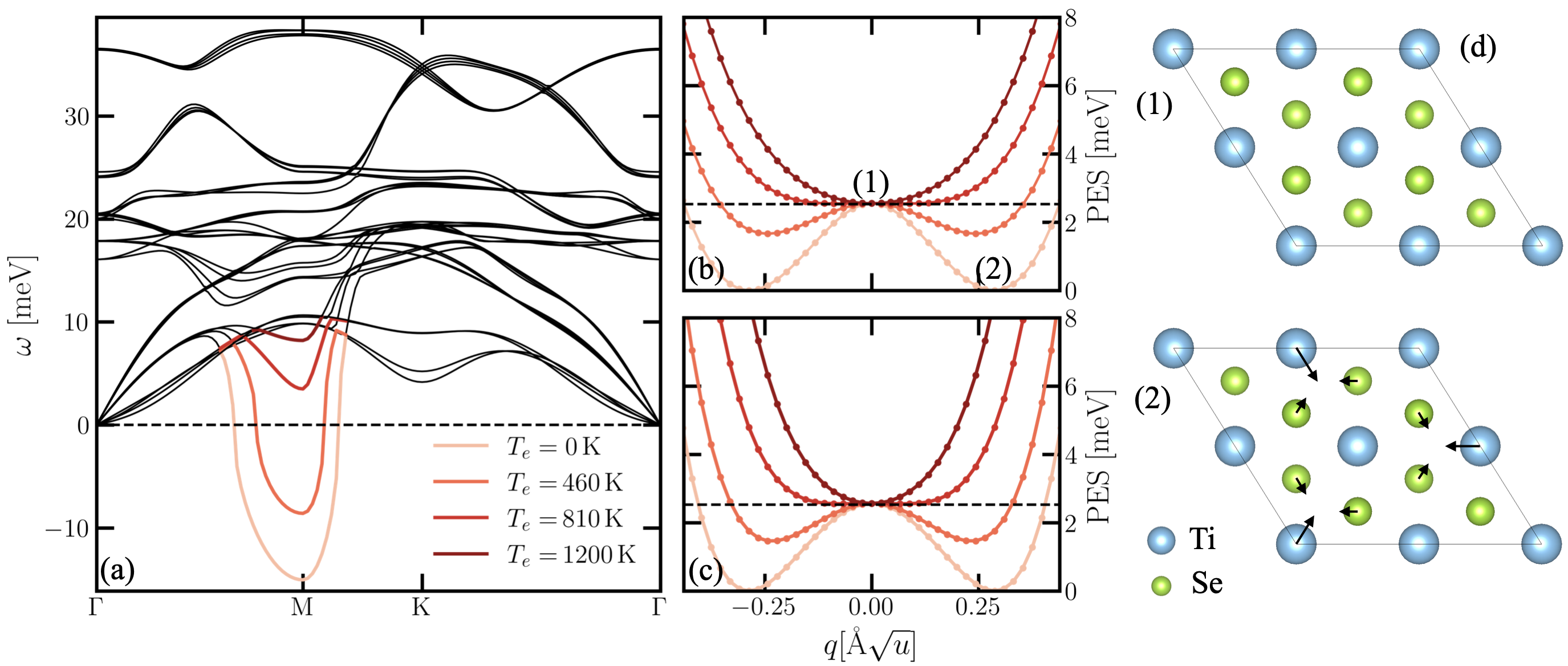}
\caption{
(a) Phonon dispersion of monolayer TiSe$_2$ along the $\Gamma$-M-K-$\Gamma$
high-symmetry path for electronic temperatures ranging between 0 and 1200~K.
The soft phonon responsible for the CDW structural distortion is color coded.
(b) Potential energy surface for mass-weighted structural displacements $q_\nu$
along the soft mode for the same electronic
temperature of panel (a).  (c) Same as (b) as obtained from explicit DFT
simulations.  (d) Crystal structures of monolayer  TiSe$_2$ in its
high-symmetry (top) and CDW phases (bottom). The atomic displacement in
presence of the CDW distortion are marked by arrows.  The high-symmetry and CDW
phases correspond to the local maximum and minima of the double-well potential,
marked by (1) and (2), respectively.  } \label{fig:PES}
\end{figure*}

For dynamically stable modes, the damping coefficient $\gamma_{\bq\nu}$ can be
expressed in terms of the phonon self-energy due to the electron-phonon and
phonon-phonon interactions \cite{Pan2025,Stefanucci2025}, providing a framework
suitable for first-principles calculations. The approach formulated in
Refs.~\cite{Pan2025,Stefanucci2025}, however, assumes dynamically stable
phonons, and it is not directly applicable to model the damped dynamics of
unstable modes across a phase transitions.  We thus interpret $\gamma_{\bq\nu}$
as a phenomenological damping parameter. 

\section{Results}\label{sec:tise2}

We apply the framework derived in Sec.~\ref{sec:1p} to model the light-induced
melting of 2$\times$2 CDW order in monolayer TiSe$_2$. All computational details are reported in the Supplemental Material (SM) \cite{supp}.  
We begin by considering the fictitious case of a static excited carrier density 
and discussing its influence on the PES.  In this limiting case, the
effects of photoexcitation on the electronic structure is approximately
accounted for by constraining electronic occupations at the Fermi level via  a
Fermi-Dirac function $f_{n\bk}(T_e) = \{\exp[ (\varepsilon_{n\bk}-\mu)/k_{\rm
B}T_e] +1\}^{-1}$ with effective electronic temperatures $T_e=$0, 460, 810, and
1200~K.  These values correspond to increases of electronic energy $\Delta E =
\sum_{n\bk} \varepsilon_{n\bk} [f_{n\bk}(T_e) - f_{n\bk}(0)]$ of 0,
6, 23, and 58~meV/uc, which are compatible with experimentally
accessible photoexcitation conditions \cite{fragkos2026electron}. 

The phonon dispersion of monolayer TiSe$_2$ obtained from DFPT in the
primitive unit cell of the high-symmetry phase is illustrated in
Fig.~\ref{fig:PES}~(a) for different electronic temperatures. 
The imaginary phonon frequency at the M
high-symmetry point reflects the dynamical instability of the crystal
towards the formation of a CDW distortion.  %
Above a critical electronic temperature of $T_e^{\rm c} \sim ${750}~K, the
imaginary phonon frequencies are replaced by real-valued, 
positive-definite frequencies, marking the phase transitions from 
the CDW to the high-symmetry phase. Away from the M point,  
the phonon dispersion is unaffected by electronic temperature.
The value of $T_e^{\rm c}$ agrees well 
with earlier theoretical works
\cite{PhysRevB.92.245131,PhysRevB.95.245136}. 
This quantity should not be confused with the critical temperature $T^{\rm c}$
for the CDW phase transition at thermal equilibrium.  The latter further
requires to account for vibrational contribution to the free energy, leading
to $T^{\rm c} < T_e^{\rm c}$.
The modifications of the PES due to a static photoexcited carrier density are
illustrated in Fig.~\ref{fig:PES}~(b), where the PES is obtained from DFT total 
energy calculations by
displacing the structure along the soft phonon mode in a 2$\times$2 supercell
for the same electronic temperatures $T_e$ of Fig.~\ref{fig:PES}~(a).  For low
$T_e$, the PES takes the form of a double well potential.  The minima and
maximum correspond to the CDW and high-symmetry phases, respectively,
illustrated in Fig.~\ref{fig:PES}~(d). Above the critical temperature $T_e^{\rm
c}$, the PES exhibits a single minimum centered at $q=0$, reflecting the
stabilization of the high-symmetry phase. 
The changes of the PES induced by photoexcited carriers can be approximately
expressed in terms of the electron-phonon interaction.  For a static
modification of electronic occupations $\Delta f_{n\bk}$  the ensuing change of
the PES is approximately given by: $ \Delta V_{\bq\nu} = \bra{\Psi_s} \hat
H_{\rm eph} \ket{\Psi_s} =  \xi_{\bq\nu} q_{\bq\nu}^2$.  The second equality
directly follows from the expectation value of the electron-phonon coupling
Hamiltonian.  The parameter $\xi_{\bq\nu}$ -- defined in Eq.~\ref{eq:EOM3} --
depends linearly on the photoexcitation density $\Delta f_{n\bk}$.  As $\Delta
f_{n\bk}$ and $ \xi_{\bq\nu}$ increase, the PES modifications introduced by $
\Delta V_{\bq\nu}$ progressively suppress the double-well character of the PES
and, above a critical value $ \xi_c$, the transition from double-well to
single-well occurs. 
Only nonlinear electron-phonon interactions contribute to $\Delta V_{\bq\nu}$,
whereas first-order electron-phonon matrix elements vanish identically due to
symmetry as discussed in Appendix~\ref{sec:sym}.  In Fig.~\ref{fig:PES}~(c), we
report the model PES obtained by adding $ \Delta V_{\bq\nu}$ to the quartic
potential defined in Eq.~\eqref{eq:tmpx} for the same electronic temperature as
in Fig.~\ref{fig:PES}~(b).  The PES agrees well with explicit DFT calculations.
{We attribute residual discrepancies to sixth- and higher-order
contributions to the ground-state PES in Eq.~\eqref{eq:tmpx} and higher-order electron-phonon interactions, as discussed in the SM.} These results demonstrate the importance of non-linear
electron-phonon interactions to capture the modifications of the PES induced by an excited carrier density. 

\begin{figure*}[t]
\includegraphics[width=0.95\linewidth]{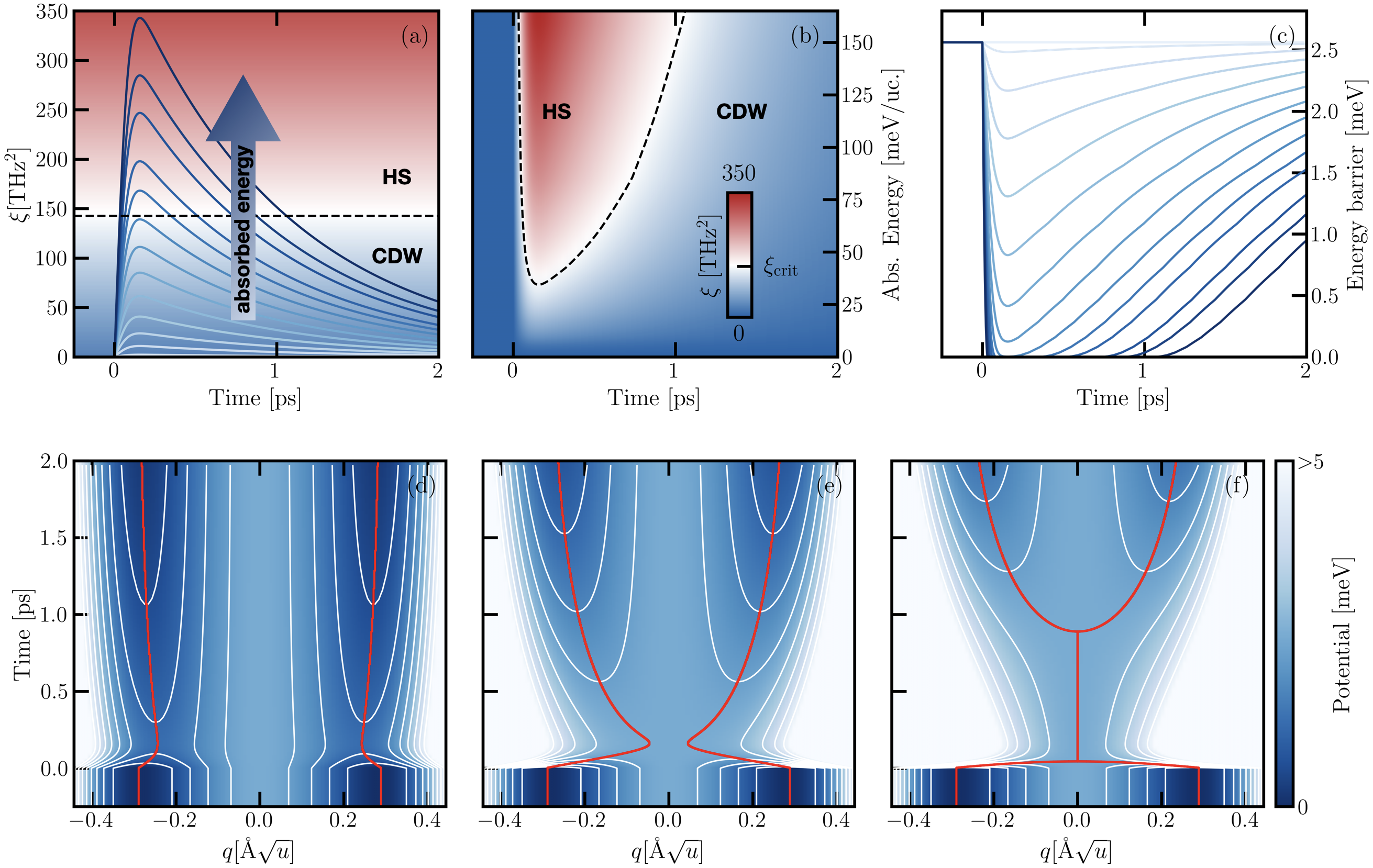}
\caption{(a) Time-dependent coupling strength $\xi$, as obtained from
Eq.~\eqref{eq:EOM3}, for absorbed energies ranging between {0} and
165~meV/uc.
The critical coupling strength $\xi_c$ required for the melting
of CDW order is marked by a dashed horizontal line.  (b) Phase diagram of CDW
melting.  Coupling strength $\xi$ as a function of time and absorbed fluence
and time.  Blue and red shading denote regions of parameter space
corresponding to the CDW and high-symmetry (HS) phase, respectively.
{(c) Time-dependent changes of the energy barrier $\Delta E$,   separating CDW-phases with opposite order parameter,   for the same absorbed of panel (a).
Time-dependent PES for displacement along the soft mode absorbed energies (d) below, (e)  near, and (f)  above the critical threshold for the melting of the CDW phase, respectively. The absolute minima of the PES are marked by solid red lines.}
}
\label{fig:xit}
\end{figure*}

\begin{figure*}[t]
\includegraphics[width=0.98\textwidth]{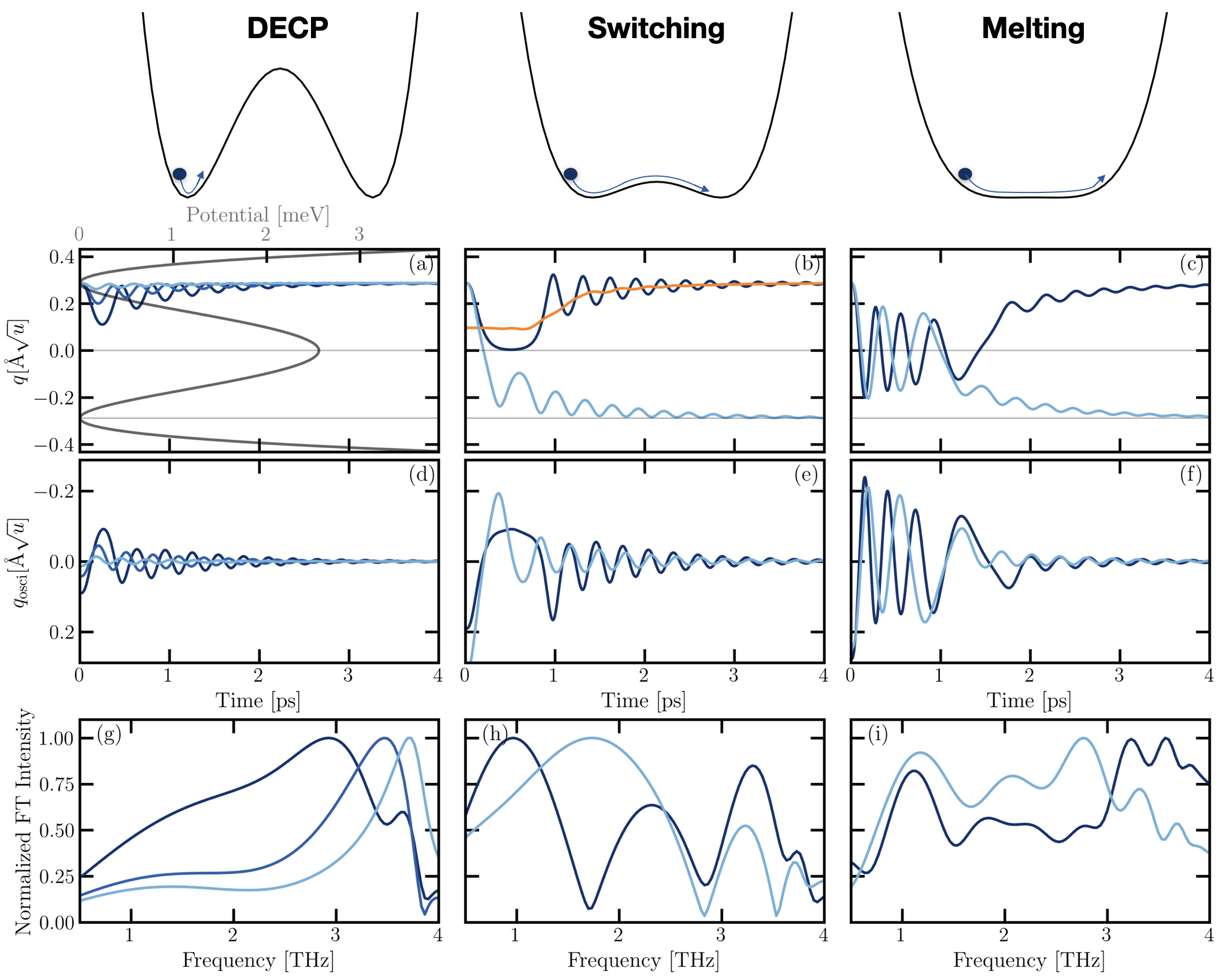}
\caption{
Real-time dynamics of the nuclear displacements for photoexcitation energies
below (a), near (b), and above (c) the critical energy threshold for symmetry
switching.  The unperturbed PES is superimposed to panel (a) with values
specified in the upper axis.  Horizontal gray lines mark minima and maximum of
the PES, corresponding to the CDW and high-symmetry phase, respectively.  (d-f)
Oscillatory component of the displacements (defined as $q_{\rm osci} = q -
q_{\rm bg}$), where $q_{\rm bg}$ are moving average backgrounds illustrated for
one displacement in orange in panel (b).  (g-i) Fourier transform of $q_{\rm
osci}$ for the same photoexcitation energies of panels (a-f).  Top: schematic
illustration of the dynamical regimes in (a-c): DECP-like structural motion
(DECP), structural switching to the degenerate CDW phase (switching), and
transient suppression of CDW order (melting). 
} \label{fig:rt_dyn}
\end{figure*}

Next, we extend these considerations to the case of a time-dependent carrier
density, $\Delta f_{n\bk} (t)$,  as induced, e.g., by absorption of a
femtosecond laser pulse and  by the ensuing carrier thermalization.  To
estimate the time-dependent changes of  $\Delta f_{n\bk} (t)$, we conducted
ultrafast dynamics simulations based on the TDBE formalism by explicitly
including the effects of electron-phonon and electron-electron interactions on
the carrier relaxation. 
The time-dependent changes of the distribution function can  alternatively be
obtained from lower-level theories -- as, e.g., the relaxation time
approximation and the  two-temperature model -- to  circumvent the cost of full
TDBE simulations.  The time-dependent coupling strength $\xi(t)$, obtained from
Eq.~\eqref{eq:EOM3} and \eqref{eq:xi_SC}, is illustrated in Fig.~\ref{fig:xit}~(a) for several
absorbed energies.  The critical coupling $\xi_c$ is marked by a dashed
horizontal line.  The strong rise of the coupling parameter $\xi$ reflects the
increase of electronic energy due to the absorption of a pump pulse, 
whereas its decrease arises from the carrier thermalization and the energy transfer to
lattice degrees of freedom due to electron-phonon coupling. 
For sufficiently large absorbed energies, the coupling strength $\xi$ exceeds
the critical threshold $\xi_c$ for the melting of CDW order.  Correspondingly,
for a finite time interval the double-well character of the PES is suppressed
and replaced by a single minimum corresponding to the high-symmetry structure.
Knowledge of the time-dependent coupling strength $\xi (t)$ enables to reconstruct the
time-dependent phase diagram for the melting of CDW order --  illustrated in
Fig.~\ref{fig:xit}~(b) -- 
as well as the transient changes of the PES induced by photoexcitation. The potential energy barrier separating CDW-phases with opposite order parameters, for example, can be expressed as $\Delta E (t) = \beta_{\bq\nu}^{-1} [|\w_{\bq\nu}|^2/2 - \xi(t) ]^2\theta(\xi_c-\xi(t))$ and it is illustrated
in Fig.~\ref{fig:xit}~(c) for the same absorbed energies as in Fig.~\ref{fig:xit}~(a). Vanishing of $\Delta E (t)$ marks the transient recovery of the high-symmetry state. 
{The time-dependent modifications of the PES along the soft mode
are further illustrated in Figs.~\ref{fig:xit}~(d-f) for absorbed energies 
 below, near, and above the critical threshold.
Before  excitation ($t<0$), the PES remains in its ground state, characterized by two distinct  minima. 
For  absorbed energies below  threshold (Fig.~\ref{fig:xit}~(d))
the PES preserves a double-well character at all times. For higher absorbed energies (Fig.~\ref{fig:xit}~(e-f)), 
the PES  assumes a single minimum for a finite interval of time, eventually recovering a double-well form after carrier thermalization.
}

\begin{figure*}[t]
\includegraphics[width=0.98\textwidth]{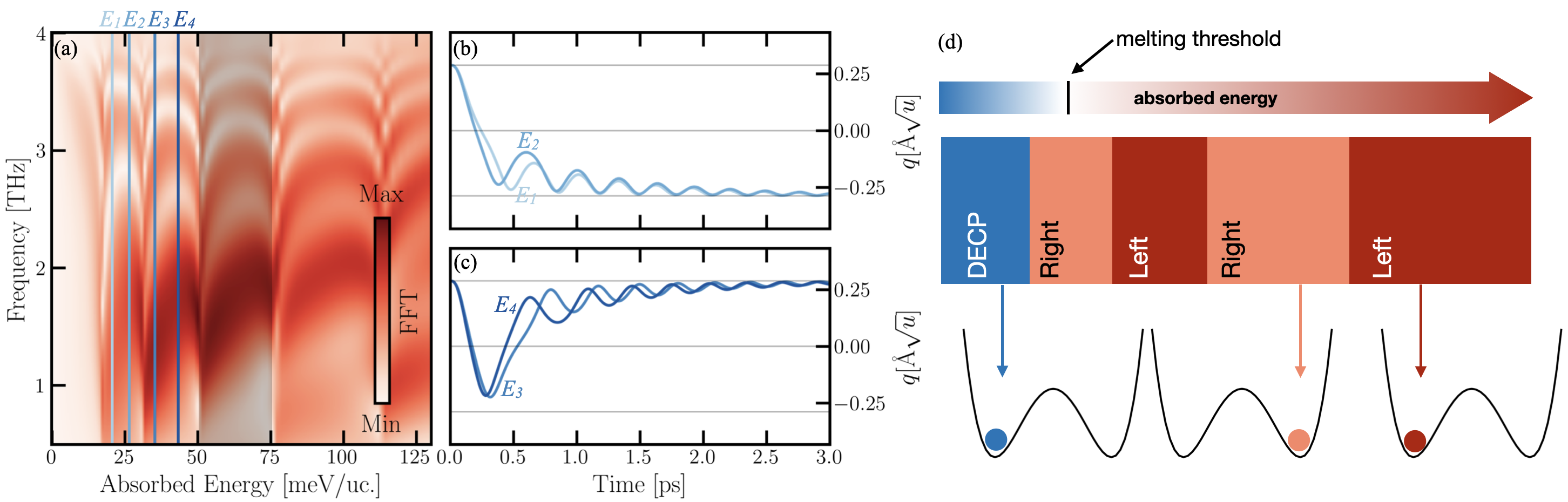}
\caption{
(a) Fourier transform of the nuclear trajectories as a function of absorbed
energy (fluence) and frequency.  Vertical lines mark the absorbed energies of
the nuclear trajectories presented in panels (b) and (c). The rectangular
shading delimit the interval of energies for which the nuclei relax in the
right minimum of the PES.  (b) Nuclear trajectories corresponding to the
absorbed energies $E_1$ and $E_2$, marked by solid lines in panel (a),
illustrating nuclear relaxation in the right minimum.  Horizontal gray lines
mark minima and maximum of the PES, corresponding to the CDW and
high-symmetry phase, respectively.  (c) Same as (b) for energies $E_3$ and
$E_4$, showing relaxation in the left minimum. (d) Schematics illustration
of the dynamical behaviour of CDW melting as a function of absorbed energy.
For low energies, the nuclear dynamics is confined to a single minimum of the
double well potential, and it is well described by the DECP approach.  For
larger energies, nuclear trajectories exhibit alternating domains, in which
the structure relax at either the left or right minima of the PES. 
}
\label{fig:FFT_spec}
\end{figure*}

We investigate the melting of CDW order in real time through solution of the
equation of motion in  Eqs.~\eqref{eq:EOM1}-\eqref{eq:EOM3}.
Figures~\ref{fig:rt_dyn}~(a-c) show the displacements along the soft phonon for
absorbed energies below, near, and above the melting threshold. Horizontal lines mark the displacements corresponding the
high-symmetry (center line) and CDW phases (top and bottom lines).  At low
excitation energies (Figures~\ref{fig:rt_dyn}~(a)), the lattice undergoes
damped oscillations in the vicinity of the local minimum. The crystal does not
abandon the CDW phase and there is no switch to the higher-symmetry phase.  In
this limit,  the dynamics is well described within the harmonic approximation,
and the  damped structural oscillations follow the characteristic behaviour
expected for the displacive excitation of coherent phonons (DECP).  For higher
absorbed energies (Fig.~\ref{fig:rt_dyn}~(b)), the structural dynamics is
strongly influenced by anharmonic effects. Nuclear trajectories can undergo
switching to CDW phase with opposite order parameter.  By further increasing the absorbed energy
(Figures~\ref{fig:rt_dyn}~(c)), the crystal lattice undergoes oscillations
around the high-symmetry phase.  This behaviour indicates the transient
recovery of a higher-symmetry state and it is the main fingerprint for the
melting of CDW order.  On longer timescales, electrons thermalize and
structural motions is damped, resulting in the recovery of the CDW phase. 
To inspect the dynamical fingerprint of CDW melting in frequency domain, we
report in Figs.~\ref{fig:rt_dyn}~(d-f) the oscillatory behaviour of the nuclear
trajectories after background subtraction, and in panels (g-i) their Fourier
transform.  At low excitation densities (Fig.~\ref{fig:rt_dyn}~(g)), the
Fourier spectrum is peaked at $\w = 3.7$~THz,  which agrees with the
frequency of the soft mode in the CDW phase.  Upon increasing the excitation
energy, this peak progressively shifts to lower frequencies, following the characteristic behaviour expected in proximity of a structural instability (see also Fig.~\ref{fig:dw}~(b)).
For larger excitation energies [Figs.~\ref{fig:rt_dyn}~(h-i)], the spectrum
develops a richer structure, revealing the coexistence of several
frequency components.

In Fig.~\ref{fig:FFT_spec}~(a), the frequency spectrum  -- obtained from
the Fourier transform of background-subtracted nuclear trajectories -- is
reported for absorbed energies up to 130~meV per unit cell. 
Below 20~meV, 
the spectrum is dominated by a single frequency. For energies above $20$~meV, these features are replaced 
by a broad frequency spectrum.  
This behaviour  qualitatively matches the results of fluence-dependent 
pump-probe optical experiments for bulk TiSe$_2$ from Ref.~\cite{PhysRevLett.107.036403}. 
The frequency spectrum exhibits distinct energy domains
characterized by the alternation of frequency softening and hardening.  One
such domain is highlighted by the shaded rectangle in Fig.~\ref{fig:FFT_spec}~(a). 
To understand the origin of these features, we report in Fig.~\ref{fig:FFT_spec}~(b-c) the nuclear trajectories for four
absorbed energies, labeled $E_{1},\dots, E_4$ in Fig.~\ref{fig:FFT_spec}~(a).
For energies $E_{1,2}$ and $E_{3,4}$, corresponding to distinct domains, the damping of the nuclear motion drives
the system into the left and right minima of the double-well potential,
respectively, indicating  condensation of CDW phases with opposite order parameters. In short, these results indicate that analyzsis of the fluence-dependent frequency spectrum of coherent structural motion, as obtained, e.g., 
in pump-probe experiments,  may provide insight into the sign of the order parameter during CDW condensation after photoexcitation.
This scenario is schematically illustrated in Fig.~\ref{fig:FFT_spec}~(d).
Supplementary Fig.~S2 illustrates that stochastic forces occurring at finite temperatures tend to suppress this effect randomizing the order parameter achieved throughout condensation of the CDW phase.

\section{Discussion}\label{sec:disc}

The formalism introduced in Sec.~\ref{sec:1p} enables the description of electron-phonon interactions and 
coherent structural motion in crystals characterized by dynamically
unstable phonons. While the discussion in Sec.~\ref{sec:tise2} has been
confined to a CDW crystal, these
concepts can equally be applied to light-induced structural motion and
phase transitions across a broad range of material classes.
As one example, perovskite crystals exhibit a degree of structural
polymorphism arising from dynamical instabilities closely reminiscent of
those found in CDW systems \cite{PhysRevB.101.155137}. Similarly, ferroelectric materials are
prototypical compounds in which the ferroelectric transition is governed by
a highly anharmonic PES, with states of opposite
ferroelectric polarization correspond to the minima of a double-well
potential, 
while the paraelectric state corresponds to the center of the potential
well.
More generally, we anticipate that the present framework can be extended to
account for more complex forms of dynamical instabilities, such as those
occurring in Kagome metals \cite{Xiao2023} and in the type-II Weyl 
semimetals WTe$_2$  and
MoTe$_2$ \cite{PhysRevLett.128.015702,horstmann2025coherentphononcontrolamplitude}. In these systems, the PES cannot be described by a simple
double-well potential. Realistic PES models must account for third-order
anharmonicities as well as the coexistence of multiple order parameters.

Several further advancements can be envisioned to extend the capabilities of the
theoretical framework presented here. In principle, finite temperature and 
quantum nuclear effects
can be incorporated into the equations of motion
(Eqs.~\eqref{eq:EOM1}–\eqref{eq:EOM3}) through the explicit inclusion of correlation 
effects in the expectation values of second- and higher-order products of operators. 
A  difficulty in this context, however, is
that anharmonic terms give rise to non-Bose–Einstein statistics,
thereby requiring an explicit evaluation of the partition function.

Finally, we note that analogous principles may be applied to model other forms
of ultrafast perturbations, including photoexcitation via multiple pulses,
direct lattice excitation via resonant THz radiation or inelastic stimulated
Raman scattering.  Taken together, these points outline several promising and
complementary directions for advancing the theoretical and computational
modelling of light-induced phase transitions.

\section{Conclusions}\label{sec:conc}

In conclusion, we developed a theoretical framework to describe the structural
dynamics induced by photoexcitation in crystal structures hosting
dynamically unstable modes.  Our approach handles quartic anharmonicities 
by combining (i) a unitary transformation of the
lattice and electron-phonon coupling Hamiltonians with (ii) the Heisenberg
equation of motion for the nuclear displacements.  The key result of our work
is an equation of motion,  suitable for first-principles calculations,  for the
structural dynamics across a light-induced phase transitions. 
We implemented this approach and applied it to describe the light-induced melting of CDW order 
in monolayer TiSe$_2$ from first principles. 
The resulting structural dynamics captures the critical features of the lattice
response across the phase transition. In particular, our simulations reproduce
the pronounced softening of the relevant modes upon approaching the critical
fluence, and they reveal the emergence of a rich, multi-component frequency
spectrum at supercritical fluences. Overall, the framework provides a
first-principles route to predictively investigate light-induced CDW melting. 

Finally, our work establishes general principles for describing
light-induced structural motion in the presence of dynamical instabilities.
These concepts provide a foundation for predictive, atomistic models of
nonequilibrium dynamics in materials well beyond the family of charge-density-wave
crystals. In particular, we anticipate that the same framework can be extended
to other material families characterized by unstable lattice modes, including
ferroics and multiferroics, as well as crystals in which long-range disorder is
intrinsic -- such as polymorphous perovskites.

\section*{acknowledgement}
This project received funding from the Deutsche Forschungsgemeinschaft (DFG
project numbers 443988403 and 499426961) and from the European Union as part of
the MSCA Doctoral Network TIMES (Grant Agreement No. 101118915).  We gratefully
acknowledge the computing time provided by the high-performance computer
Lichtenberg at the NHR Centers NHR4CES at TU Darmstadt (Project p0021280).
Fruitful discussions with Laurenz Rettig, Yiming Pan, and Joris Kronenberg are gratefully acknowledged.

\appendix

\section{Equation of motion for symmetry switching in a one-dimensional model}\label{sec:1Dmodelder}

We determine the equation
of motion for a particle in the shifted quartic potential 
in Heisenberg picture.
We start from the general formula, which follows directly
from the Heisenberg equation of motion (cf.~Eq.~(17) 
in Ref.~\cite{Caruso2023}): 
\begin{align} 
\partial_t^2 Q + \Omega^2 Q = - \frac{\Omega}{\hbar} 
\langle [\hat P, \hat V^{(a)}+ \Delta \hat V(t)] \rangle \quad. 
\end{align}
where $\hat P = \hat a - \hat a^\dagger$ and 
$\langle\dots\rangle $ denotes the ground-state expectation value. 
$\hat V^{(a)}$ and $\Delta \hat V(t)$ are the anharmonic term
and the time-dependent potential of the transformed Hamiltonian, 
Eq.~\eqref{eq:H2}. 
All commutators can be evaluated via application of the general identity: 
\begin{align}  
[\hat P,\hat Q^n] =  2n \hat Q^{n-1}
\quad. 
\end{align}
Evaluation of the commutator yields:

\begin{align} 
- \frac{\Omega }{\hbar}
[\hat P, \hat V^{(a)}] &= - \frac{3}{2} \sqrt{{\hbar \Omega \beta}} Q^2
-
\frac{\hbar \beta}{2\Omega} Q^3 
\\ 
-\frac{\Omega}{\hbar}[\hat P, \Delta \hat V(t)] &= 
-
2\xi (t)
\left [   \hat Q  + 
\left(\frac{{\Omega^3}}{{\hbar\beta}} \right)^{1/2}
\right]
\end{align}

Combining all terms, we obtain the following equation of motion 
for the coordinate: 
\begin{align}
\partial_t^2 Q + \gamma \partial_t Q+ \Omega^2 Q = 
D(t)
\label{eq:HO1}
\end{align}
where: 
\begin{align}
D(t) = - \frac{3}{2} \sqrt{{\hbar \Omega \beta}} Q^2
- \frac{\hbar \beta}{2\Omega} Q^3  
-
2\xi (t)
\left [   \hat Q  +
\left(\frac{{\Omega^3}}{{\hbar\beta}} \right)^{1/2}
\right]
\label{eq:D1}
\end{align}

We introduced a damping rate $\gamma$, which is justified according to the procedure discussed in Ref.~\cite{Pan2025}. 
Equations~\eqref{eq:HO1} and \eqref{eq:D1} recast the equation of 
motion in a quartic potential in the form of a driven, damped harmonic 
oscillator. Anharmonic effects, however, are fully retained and 
incorporated in the driving force $D$. 
In the limit of small oscillations, one can retain only 
the zero-th order term in the driving force, Eq.~\eqref{eq:D1}. 
In this case, the motion reduces to a simple driven oscillator, and 
the equation of motion is equivalent to the case of the displacive 
excitation of coherent phonons. 

\section{Electron-phonon interaction Hamiltonian for dynamically unstable modes}\label{sec:AHep}
In the following, 
we derive the second-quantized form of 
the EPI Hamiltonian in presence of dynamically unstable modes in the specific 
case of a quartic double-well potential along a symmetry-breaking phonon mode. 
We start from the normal coordinate representation of the first- and second-order EPI Hamiltonians \cite{RevModPhys.89.015003,StefanucciPRX}: 
\begin{align}
\hat H^{(1)}_{\rm eph} 
\label{eq:htmp1}
&= 
\frac{1}{\sqrt{N_p} } 
\sum_{\bq\nu} 
\sum_{nm\bk} 
\bra{\psi_{m\bkq}} \partial_{\bq\nu} v_{\rm KS} \ket{\psi_{n\bk}}
\\ 
&\quad 
\quad
\quad\quad
\times [ \hat c_{m\bkq}^\dagger \hat c_{n\bk} - f_{n\bk}^0 \delta_{nm} \delta_{\bq,0} ] 
\hat q_{\bq\nu} \quad, \nonumber\\
\hat H^{(2)}_{\rm eph}&=  
\label{eq:htmp2}
\frac{1}{N_p }
\sum_{\bq\nu}
\sum_{\bq'\nu'}
\sum_{nm\bk}
\bra{\psi_{m\bkq+\bq'}} 
\partial_{\bq'\nu'} 
\partial_{\bq\nu} 
v_{\rm KS} \ket{\psi_{n\bk}} 
\nonumber
\\ 
&\times [  \hat c_{m\bkq+\bq'}^\dagger \hat c_{n\bk}  
-f_{n\bk}^0 \delta_{nm}\delta_{\bq+\bq',0}] 
\hat q_{\bq\nu}
\hat q_{\bq'\nu'}
\quad. 
\end{align}
$\partial_{\bq\nu} v_{\rm KS}$  
is the first-order variation of the 
Kohn-Sham potential for atomic displacements along the normal mode 
$\bq\nu$, and $\partial_{\bq\nu}  \partial_{\bq' \nu'} v_{\rm KS}$ 
is the second-order variation. They are 
defined by: 
\begin{align}
\partial_{\bq\nu} v_{\rm KS}  
&=  \sum_{\k p}  \left.\frac{\D v_{\mathrm{KS}} (\br,\{\boldsymbol{\tau}\})}{\D
{\Delta} {\tau}_{\k p \a} }\right|_0 e^{i\bq{\bf R}_p} \frac{{ e}^{\k\a}_{\bq\nu}}{\sqrt {M_\k}} \quad, 
\\
 \partial_{\bq' \nu'} \partial_{\bq\nu} v_{\rm KS} 
&= \frac{1}{2} \sum_{\k p \a } \sum_{\k'p' \a'}  \left.
 \frac{\D^2 v_{\mathrm{KS}} (\br,\{\boldsymbol{\tau}\})}{\D
{\Delta} {\tau}_{\k' p'\a'} \D
{\Delta} {\tau}_{\k p \a} 
}
\right|_0 
\\ 
&
\quad\quad
\times e^{i(\bq{\bf R}_p + \bq'{\bf R}_{p'}) }
\frac{{ e}^{\k'\a'}_{\bq'\nu'}{ e}^{\k\a}_{\bq\nu}}{\sqrt {M_{\k'} 
M_\k}}\quad.
\nonumber
\end{align}
As detailed in Refs.~\cite{Marini2015,StefanucciPRX}, the appearance of the equilibrium distribution function $f_{n\bk}^0$ follows 
from the requirement that momenta and their derivatives vanish at equilibrium. 
The forms for the first- and second-order EPI Hamiltonians specified by 
Eqs.~\eqref{eq:htmp1} and \eqref{eq:htmp2} 
are valid regardless of whether the normal mode $\bq\nu$ 
is dynamically stable or not.  
Following the same procedure outlined for the phonon Hamiltonian, 
we introduce the second-quantized  form for bosonic operators 
by distinguishing between stable ($\mathcal S$) and unstable 
($\mathcal U$) normal modes, and by restricting the discussion to anharmonicities 
in the form  double-well potentials. We thus partition phonon sums according 
to $\sum_{\bq\nu} = \sum_{\bq\nu}^\mathcal U + \sum_{\bq\nu}^\mathcal S$, 
leading to the following partioning 
of the first- and second-order EPI Hamiltonians: 
\begin{align} 
\hat H^{(1)}_{\rm eph}  &= 
\hat H^{(1)\,\mathcal S}_{\rm eph}  
+ \hat H^{(1)\,\mathcal U}_{\rm eph}   
\\ 
\hat H^{(2)}_{\rm eph} &= 
\hat H^{(2)\,\mathcal {SS}}_{\rm eph}
+ \hat H^{(2)\,\mathcal {US}}_{\rm eph}
+ \hat H^{(2)\,\mathcal {UU}}_{\rm eph}
\label{eq:Hfirst}
\end{align} 
For stable modes, bosonic operators can be rewritten in second
quantization via  Eq.~\eqref{eq:2qQ}. 
Substitution in Eqs.~\eqref{eq:htmp1} and \eqref{eq:htmp2} one recovers the
usual second-quantized form of the first- and second-order EPI Hamiltonian: 
\begin{align}
\hat H^{(1) \,\mathcal S}_{\rm eph} &= 
\frac{1}{\sqrt{N_p} }
\sum_{\bq\nu}^{\mathcal S}
\sum_{nm\bk} g _{nm}^\nu(\bk,\bq) 
\\&
\quad\quad
\times
[ \hat c_{m\bkq}^\dagger \hat c_{n\bk} - f^0_{n\bk} \delta_{nm} \delta_{\bq,0} ] 
\hat Q_{\bq\nu}
\quad,\nonumber
\\ 
\hat H^{(2)\,\mathcal {SS}}_{\rm eph}&= 
\frac{1}{N_p } \sum_{\bq\nu}^\mathcal {S}
\sum_{\bq'\nu'}^\mathcal {S} \sum_{nm\bk}  
g _{nm}^{\nu\nu'}(\bk,\bq,\bq')  
\\&
 [ \hat c_{m\bkq+\bq'}^\dagger \hat c_{n\bk} - f^0_{n\bk} \delta_{nm} \delta_{\bq+\bq',0} ] 
\hat Q_{\bq\nu}\hat Q_{\bq' \nu'} 
\quad, \nonumber
\end{align} 
with $\hat Q_{\bq\nu} = \hat a_{\bq\nu} + \hat a^\dagger{-\bq\nu}$. 
We introduced the EPI matrix elements:
\begin{align}
&g _{nm}^{\nu}(\bk,\bq) = 
\bra{\psi_{m\bkq+}}
\partial_{\bq\nu}
v_{\rm KS} \ket{\psi_{n\bk}} 
\left(\frac{\hbar} {2\w_{\bq\nu}}  \right)^{\frac{1}{2}}
, 
\\
&g _{nm}^{\nu\nu'}(\bk,\bq,\bq') = 
\bra{\psi_{m\bkq+\bq'}}
\partial_{\bq'\nu'}
\partial_{\bq\nu}
v_{\rm KS} \ket{\psi_{n\bk}} 
\\&
\quad\quad\quad
\quad\quad\quad
\quad\quad\quad
\times \left(\frac{\hbar^2} {4 \w_{\bq\nu}\w_{\bq'\nu'}}  \right)^\frac{1}{2}
\quad. 
\nonumber
\end{align} 
For dynamical unstable modes, however, the second-quantized 
form for the bosonic operators specified  by Eqs.~\eqref{eq:2qQ} 
is ill-defined, since the vibrational frequency $\w_{\bq\nu}$ is 
purely imaginary. 
We circumvent this problem via the coordinate 
transformation defined in Eq.~\eqref{eq:coordtr}, 
leading to:
\begin{widetext}
\begin{align}
\hat H^{(1)\,\mathcal {U}}_{\rm eph} &=
\frac{1}{\sqrt{N_p} }
\sum_{\bq\nu}^{\mathcal U}
\sum_{nm\bk} \tilde g _{nm}^\nu(\bk,\bq) 
[ \hat c_{m\bkq}^\dagger \hat c_{n\bk} - f^0_{n\bk} \delta_{nm} \delta_{\bq,0} ]
\left[ \hat Q_{\bq\nu}   + 
\left(\frac{ \Omega_{\bq\nu}^3} {\hbar\beta_{\bq\nu}}\right)^{1/2}
\right]
\\ 
\hat H^{(2)\,\mathcal {UU}}_{\rm eph} &=  
\frac{1}{N_p } \sum_{\bq\nu}^\mathcal{U}
\sum_{\bq'\nu'}^\mathcal{U} \sum_{nm\bk}
\tilde g _{nm}^{\nu\nu'}(\bk,\bq,\bq')
 [ \hat c_{m\bkq+\bq'}^\dagger \hat c_{n\bk} - f^0_{n\bk} \delta_{nm} \delta_{\bq+\bq',0} ]
\left[\hat Q_{\bq\nu} \hat Q_{\bq' \nu'} 
+  2 
\hat Q_{\bq' \nu'}
\left(\frac{ \Omega^3_{\bq\nu}} {\hbar \beta_{\bq\nu}}\right)^{1/2} 
 \right] 
\\ 
\label{eq:HUS}
\hat H^{(2)\,\mathcal {US}}_{\rm eph} &=
\frac{1}{N_p } \sum_{\bq\nu}^\mathcal{U}
\sum_{\bq'\nu'}^\mathcal{S} \sum_{nm\bk}
\tilde{\tilde g} _{nm}^{\nu\nu'}(\bk,\bq,\bq')
 [ \hat c_{m\bkq+\bq'}^\dagger \hat c_{n\bk} - f^0_{n\bk} \delta_{nm} \delta_{\bq+\bq',0} ]
\left[  
\hat Q_{\bq\nu}   +
\left(\frac{ \Omega_{\bq\nu}^3} {\hbar \beta_{\bq\nu}}\right)^{1/2}
\right] 
\hat Q_{\bq' \nu'} 
\end{align}
The coupling matrix elements are defined in terms of the renormalized 
frequency $\Omega_{\bq\nu}$, according to: 
\begin{align} 
\tilde g _{nm}^{\nu}(\bk,\bq) &=
\bra{\psi_{m\bkq+}}
\partial_{\bq\nu}
v_{\rm KS} \ket{\psi_{n\bk}}
\left(\frac{\hbar} {2\Omega_{\bq\nu}}  \right)^{1/2}
\\
\tilde g _{nm}^{\nu\nu'}(\bk,\bq,\bq') &=
\bra{\psi_{m\bkq+\bq'}}
\partial_{\bq'\nu'}
\partial_{\bq\nu}
v_{\rm KS} \ket{\psi_{n\bk}}
\left(\frac{\hbar^2} {4 \Omega_{\bq\nu}\Omega_{\bq'\nu'}}  \right)^{1/2}
\\
\tilde {\tilde g} _{nm}^{\nu\nu'}(\bk,\bq,\bq') &=
\bra{\psi_{m\bkq+\bq'}}
\partial_{\bq'\nu'}
\partial_{\bq\nu}
v_{\rm KS} \ket{\psi_{n\bk}}
\left(\frac{\hbar^2} {4 \Omega_{\bq\nu}\w_{\bq'\nu'}}  \right)^{1/2}
\label{eq:Hlast}
\end{align}
\end{widetext}
In summary, Eqs.~\eqref{eq:Hfirst}-\eqref{eq:Hlast} specify the 
first- and second-order EPI Hamiltonian for a system with a set 
of dynamical unstable modes characterized by quartic anharmonicities. 

\section{Coherent lattice motion with dynamical instabilities} \label{sec:AEOM}

We determine the structural dynamics induced by optical 
excitation through solution of the Heisenberg eiquation of motion for 
the position operator:
\begin{align}  
\D_t^2\hat Q_{\bq\nu} = -\hbar^{-2}
\big[[\hat Q_{\bq\nu}, \hat H], \hat H\big]\quad, 
\end{align}
where $\hat H$ denotes the full lattice Hamiltonian 
in presence of EPI:  
\begin{align} 
\hat H = \hat H_{\rm ph} + \hat H_{\rm eph}\quad. 
\end{align}
We are interested in the solution of the Heisenberg equation of 
motion for dynamically unstable modes ($\bq\nu \in \mathcal U$). 
For dynamically stable modes, the leading contribution to coherent lattice 
motion arises from the first order EPI and leads to the displacive 
excitation of coherent phonons. 
Following the same steps as in Ref.~\cite{Caruso2023}, 
the Heisenberg equation of motion leads to: 
\begin{align}
\D_t^2\hat Q_{\bq\nu} &+ \Omega_{\bq\nu} ^2 Q_{\bq\nu}   
= 
\\&
-\hbar^{-1} \Omega_{\bq\nu} 
\langle  [\hat P_{\bq\nu}, 
 \hat V^{(a)}
+ 
\hat H_{\rm eph}^{(1)} 
+ 
\hat H_{\rm eph}^{(2)} 
] 
\rangle \quad. \nonumber
\end{align}
All commutators can be evaluated via 
repeated application of the formula 
$[\hat P_{\bq\nu},  \hat Q_{\bq'\nu'}^n] = \delta_{\bq\bq'}\delta_{\nu\nu'}2n  \hat Q_{\bq\nu}^{n-1} $: 
\begin{align} 
[\hat P_{\bq\nu}, \hat V^{(a)} ] 
&= 
\frac{3}{2} \sqrt{\frac{\beta_{\bq\nu} \hbar^3}{\Omega_{\bq\nu}}}
\hat Q_{\bq\nu}^2
+ \frac{\beta_{\bq\nu}\hbar^2}{2 \Omega_{\bq\nu}^2} \hat Q_{\bq\nu}^3
\\ 
[\hat P_{\bq\nu}, \hat H^{(1)\,\mathcal U} ]
&= {2}N_p^{-1/2}
\sum_{n\bk} 
\tilde g_{nn}^\nu (\bk,0) \Delta f_{n\bk}(t)
\\  
[\hat P_{\bq\nu}, \hat H^{(1)\,\mathcal S} ]
&= 0 
\\ 
[\hat P_{\bq\nu}, \hat H^{(2)\,\mathcal {UU}}]
&= 4 N_p^{-1} \sum_{\nu'}^{\mathcal U} \sum_{n\bk}
\tilde g_{nn}^{\nu\nu'} (\bk,\bq,-\bq) 
\Delta f_{n\bk}(t) 
\nonumber
\\
&
\quad\quad 
\quad\quad 
\times\left [
 \hat Q_{-\bq\nu'} 
+  \left(\frac{\Omega_{\bq\nu}^3}{{\hbar\beta_{\bq\nu}}} \right)^{1/2}
\right]
\\
\nonumber
[\hat P_{\bq\nu}, \hat H^{(2)\,\mathcal {US}} ]
&= 
2 N_p^{-1} 
\sum_{\nu'}^{\mathcal{S}} \sum_{n\bk}
\tilde{\tilde g}_{nn}^{\nu\nu'}(\bk,\bq,-\bq)   
\\ & \quad\quad
\quad\quad
\times
\Delta f_{n\bk} (t) \hat Q_{-\bq\nu'}
\\
[\hat P_{\bq\nu}, \hat H^{(2)\,\mathcal {SS}} ]
&= 0
\end{align}
The symmetry analysis presented in Sec.~\ref{sec:sym}
indicates that the first-order matrix element $\tilde
g_{nn}^\nu{\bk,0}$  is non-zero only for phonons which
preserve the total symmetry of the lattice. 
The second-order matrix elements  $\tilde
g_{nn}^{\nu\nu'}{\bk,\bq,-\bq}$ is finite for $\nu=\nu'$, 
however, strict symmetry selection rules apply for 
off-diagonal coupling matrix element. 
In this work, we thus include $\nu=\nu'$.
Combining all terms, we obtain 
Eqs.~\eqref{eq:EOM1}-\eqref{eq:EOM3}
in the main text.

\section{Symmetry analysis of the EPI matrix element} \label{sec:sym}
The coherent phonon equation of motion involves  
the diagonal components of the first- and second-order EPI matrix elements: 
\begin{align}
\tilde g_{nn}^\nu (\bq,0) &\propto
\bra{\psi_{n\bk}}
 \partial_{\bq\nu} v_{\rm KS} 
\ket{\psi_{n\bk}}
\quad,
\\ 
{\tilde g}_{nn}^{\nu\nu'}(\bk,\bq,-\bq)  
& \propto 
\bra{\psi_{n\bk}} \partial_{\bq\nu} \partial_{\bq'\nu'}
v_{\rm KS}    \ket{\psi_{n\bk}} \quad, \\ 
\tilde{\tilde g}_{nn}^{\nu\nu'}(\bk,\bq,-\bq)
& \propto 
\bra{\psi_{n\bk}} \partial_{\bq\nu} \partial_{\bq'\nu'} 
v_{\rm KS}    \ket{\psi_{n\bk}} 
\quad.
\end{align}
A basic group-theoretical analysis enables the identification of the symmetry selection rules for the 
excitation of coherent phonons. 
Inner products in the form $\bra{A} B \ket{A}$ are non-zero onlt if  the products of the irreducible represetations 
$\Gamma_A \otimes \Gamma_B \otimes \Gamma_A$ contains the totally 
symmetric representation $\Gamma_{1}$. However, since $\Gamma_A \otimes \Gamma_A = \Gamma_1$, this condition can only be obeyed if $\Gamma_B \supseteq \Gamma_1$. 
Since the quantity $ \partial_{\bq\nu} v_{\rm KS}$ transforms like the irreducible representation of the mode $\bq\nu$, the matrix element $\tilde g_{nn}^\nu (\bq,0)$ differs from zero only for phonons $A_1$ symmetry. 
This requirements forms the basis for the symmetry selection rule for the 
excitation of coherent phonons via the DECP mechanism. 
Similar considerations can be applied to the second-order matrix elements. 
The quantity $\partial_{\bq\nu} \partial_{\bq'\nu'} v_{\rm KS}$  
transform like the product 
$\Gamma_{\bq\nu} \otimes \Gamma_{-\bq\nu'}$. 
Matrix elements of the type $\tilde{\tilde
g}_{nn}^{\nu\nu'}(\bk,\bq,-\bq)$, thus, may only differ from
zero if $\Gamma_{\bq\nu} \otimes \Gamma_{-\bq\nu'}\supseteq
\Gamma_1$. This condition is always obeyed for $\nu=\nu'$ 
however, it provides a stringent selection rule for the 
second-order coupling to other modes. 
In this work, we thus restrict ourselves to consider 
only the diagonal components $\nu=\nu'$ of the second-order 
matrix element. 

\section{Relation between the first and second-order electron-phonon coupling matrix elements} \label{sec:2to1}
In this note, we derive a
relation between the second-order EPI matrix element 
of the high-symmetry phase and first-order matrix element 
of the low-symmetry phase. 
The adiabatic force acting on the ions and arising from a 
photo-excited electron density 
is obtained by evaluating the energy derivative:  
\begin{align}
    \left.{F_{\bq\nu}}\right|_{0}  
&= -\left.\frac{\partial E^{(a)}}{ \partial Q_{\bq\nu}} \right|_{Q=0} \quad, 
\end{align}
$E^{(a)} = \bra{\psi^s_{\rm el}} \hat H_{\rm eph}
\ket{\psi^s_{\rm el}}$ is the adiabatic 
contribution to the 
lattice energy arising from the EPI in presence 
of an electronic excited $\psi^s_{\rm el}$, 
which results in the change $\Delta f_{n\bk}$ of electronic occupation $f_{n\bk}$. For dynamically unstable 
modes, the adiabatic energy is readily estimated by
substituting the fermionic operators in
Eqs.~\eqref{eq:HUU} and \eqref{eq:HUS} by $\Delta
f_{n\bk}$.  The force acting on the lattice can thus be
estimated straightforwardly by differentiation: 
\begin{align}
    \left.{F_{\bq\nu}}\right|_{Q=0}
    &=- 2 \sum_{n \bf k}
    \tilde{g}^{\nu \nu}_{nn}({\bf k},{\bf q},-{\bf q}) 
     \Delta {{f}}_{n \bf k} \left(\frac{\Omega^3_{{\bf q} \nu}}{\hbar \beta_{\nu}}\right)^{1/2} 
\end{align}
The matrix elements 
entering the definition of EPI Hamiltonian  
are relative to the high-symmetry phase. 
In the context of CDW materials, they are evaluated in the unit cell for the normal phase. 
Alternatively, the EPI Hamiltonian and 
the corresponding adiabatic energy contribution 
can be introduced starting from the low-symmetry
structure, namely for the CDW phase in its 
supercell reconstruction.
 In this case, since the low-symmetry
structure only exhibit dynamically stable mode, the 
usual form for the EPI Hamiltonian applies. 
Correspondingly, the forces acting on the 
equilibrium geometry ($Q=0$) take the form: 
\begin{align}
     \left.F_{\bq\nu}\right|_{Q=0} 
     &= - \sum_{n \bf k}
    g^{\nu}_{nn}({\bf k},0) 
     \Delta {f}_{n \bf k} 
\end{align}
Here, all indices refer to the low-symmetry structure. 
Since forces must coincide regardeless of the choice
of the unit cell, one arrives at the equality: 
\begin{align}
     2 \left(\frac{ \Omega^3_{{\bf q} \nu}}{ \hbar \beta_{\nu}}\right)^{1/2}
    \sum_{n \bf k}  
    \tilde{g}^{\nu \nu}_{nn}({\bf k},{\bf q},{\bf -q}) 
     \Delta {{f}}_{n \bf k}  
    & =
     \sum_{n \bf k}  g_{nn}({\bf k,0})
    \Delta {f}_{n \bf k}
\end{align}
All quantities on the left-hand side are evaluate
for the high-symmetry structure (i.e., in the unit
cell), whereas quantities in the right-hand side
are evaluated for the low-symmetry structure (i.e.,
in the super cell), and all momenta and band 
indices are folded.  
This identity provides a way to
replace calculations of the second-order
EPI matrix element in the unit cell, by calculations of the first-order matrix elements in the 
supercell.

\begin{align}\label{eq:xi_SC}
    \sum_{n \bf k}  
    \tilde{g}^{\nu \nu}_{nn}({\bf k},{\bf q},{\bf -q}) 
     \Delta {{f}}_{n \bf k}  
    & =
    \frac{1}{2} \left(\frac{ \hbar \beta_{\nu}}{ \Omega^3_{{\bf q}_0 \nu}}\right)^{1/2}
     \sum_{n \bf k}  g_{nn}({\bf k,0})
    \Delta {f}_{n \bf k}
\end{align}

\bibliography{references}

\clearpage

\begin{figure}[htpb]
    \centering
    \includegraphics[width=0.99\textwidth]{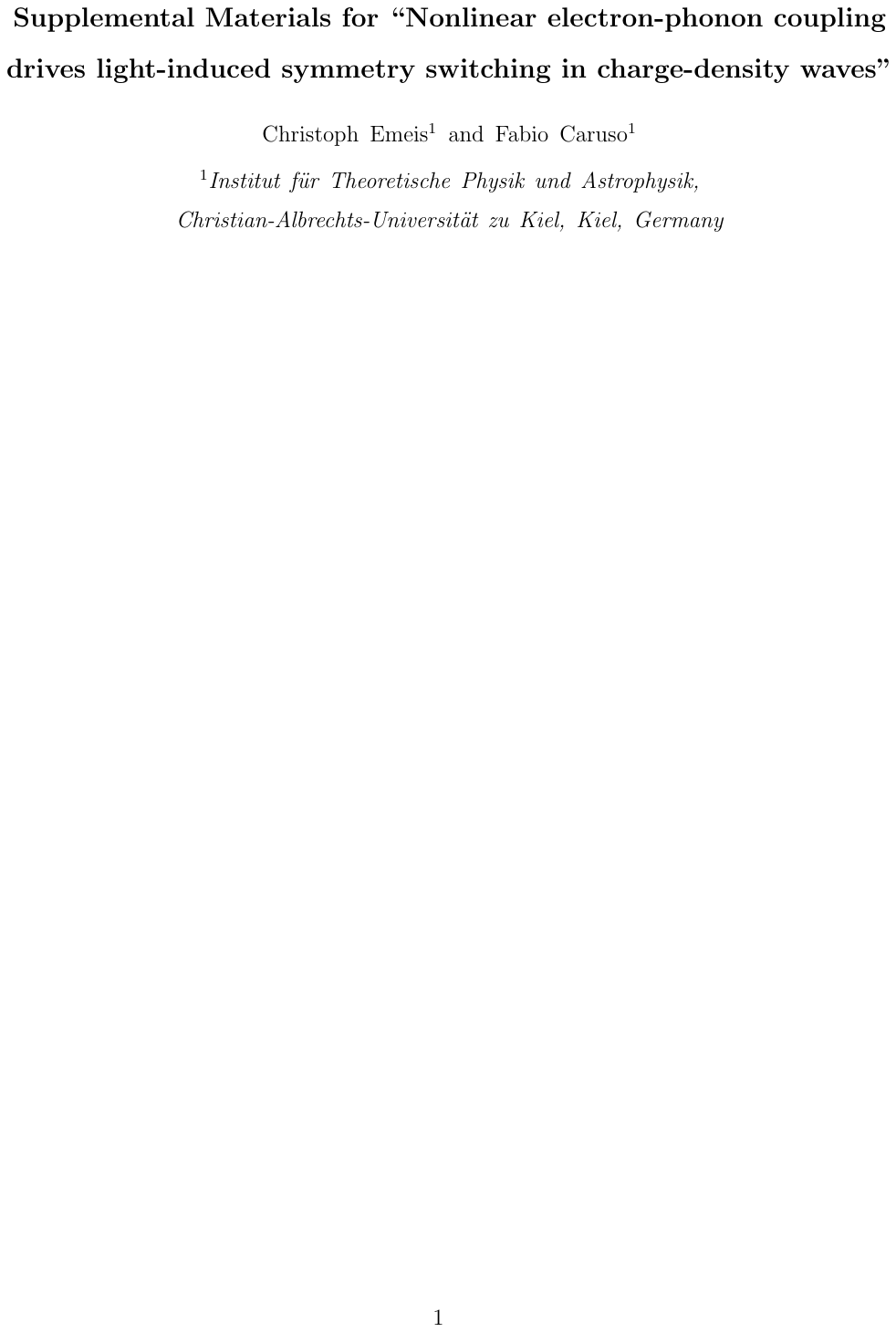}
\end{figure}

\begin{figure}[htpb]
    \centering
    \includegraphics[width=0.99\textwidth]{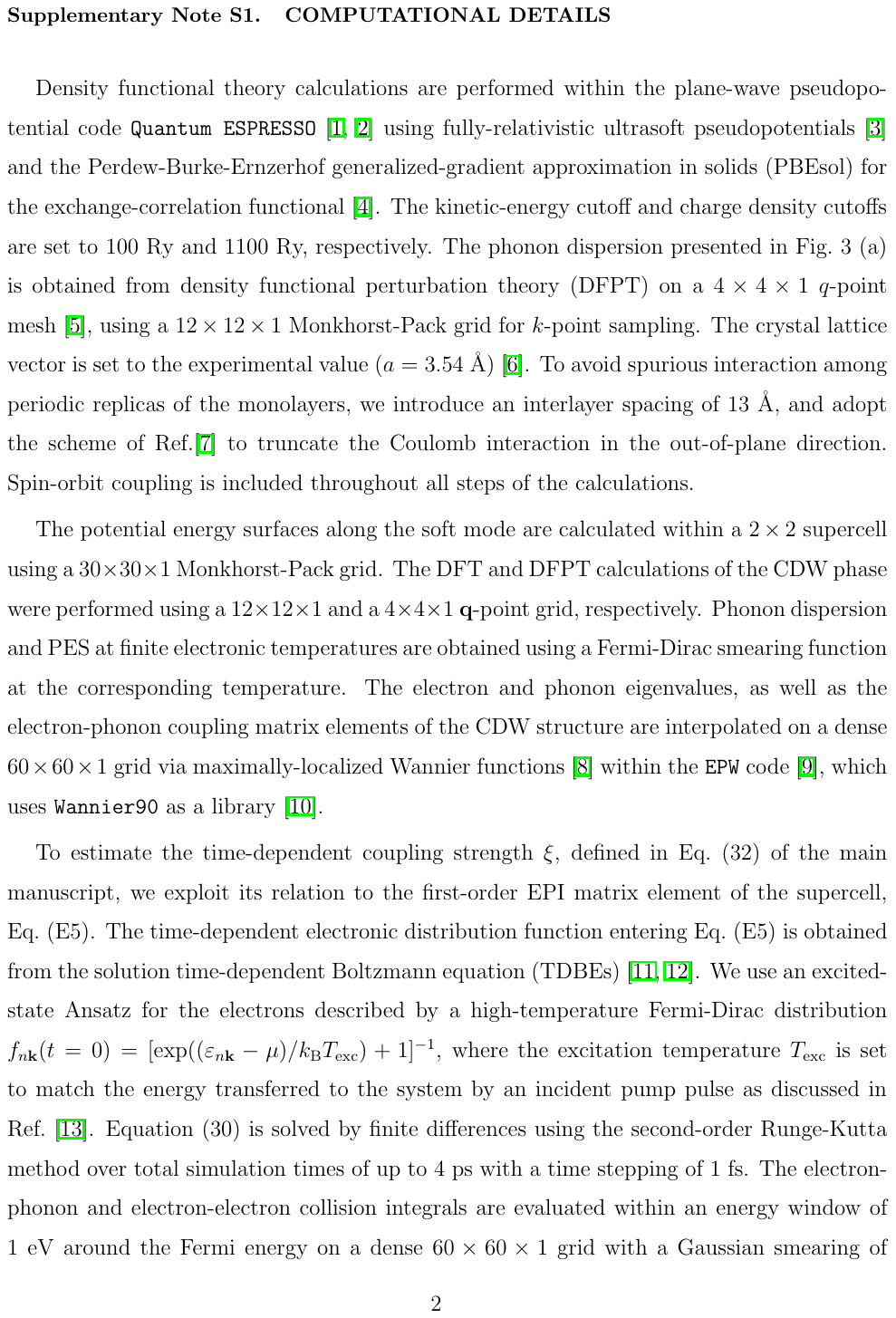}
\end{figure}

\begin{figure}[htpb]
    \centering
    \includegraphics[width=0.99\textwidth]{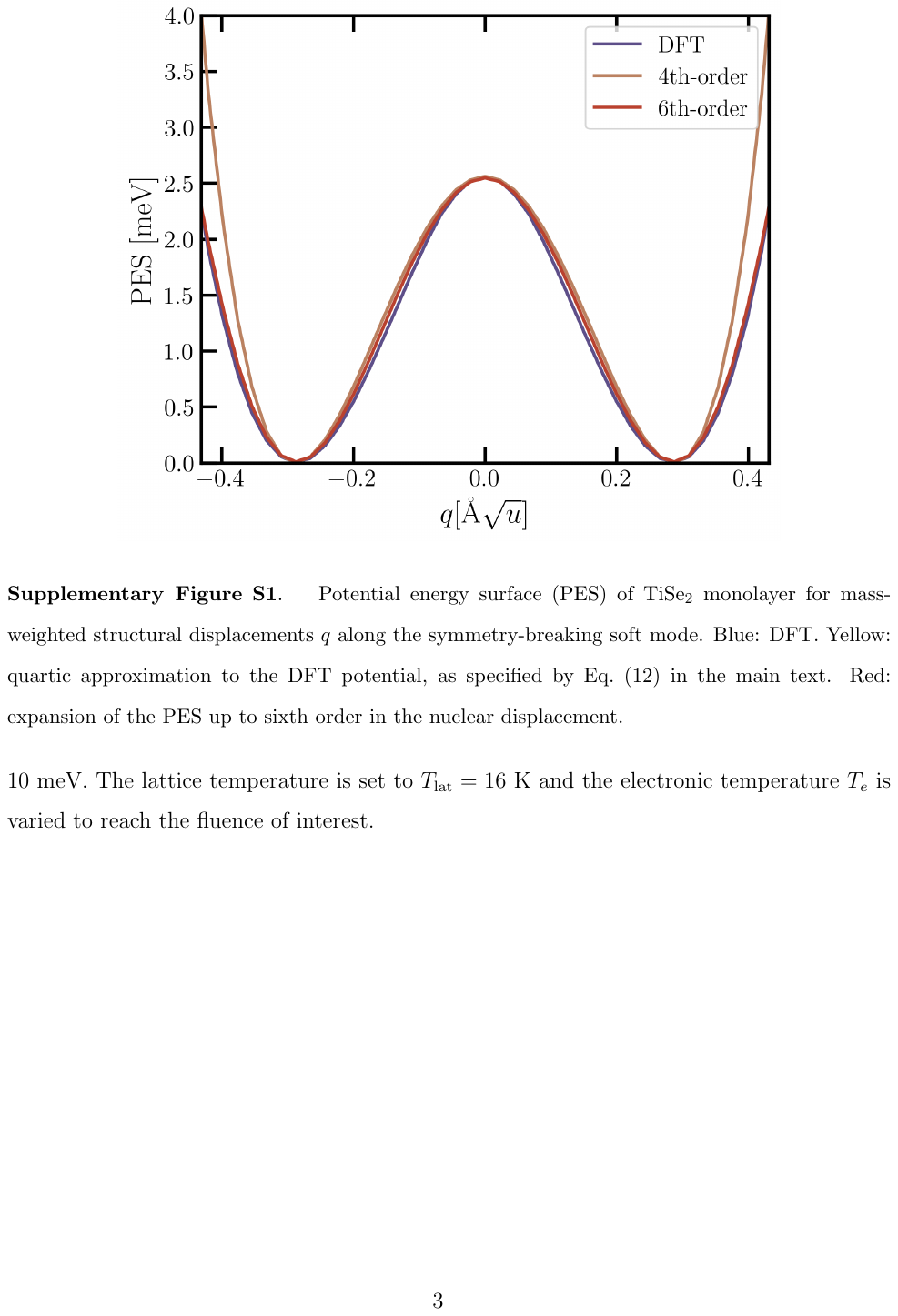}
\end{figure}

\begin{figure}[htpb]
    \centering
    \includegraphics[width=0.99\textwidth]{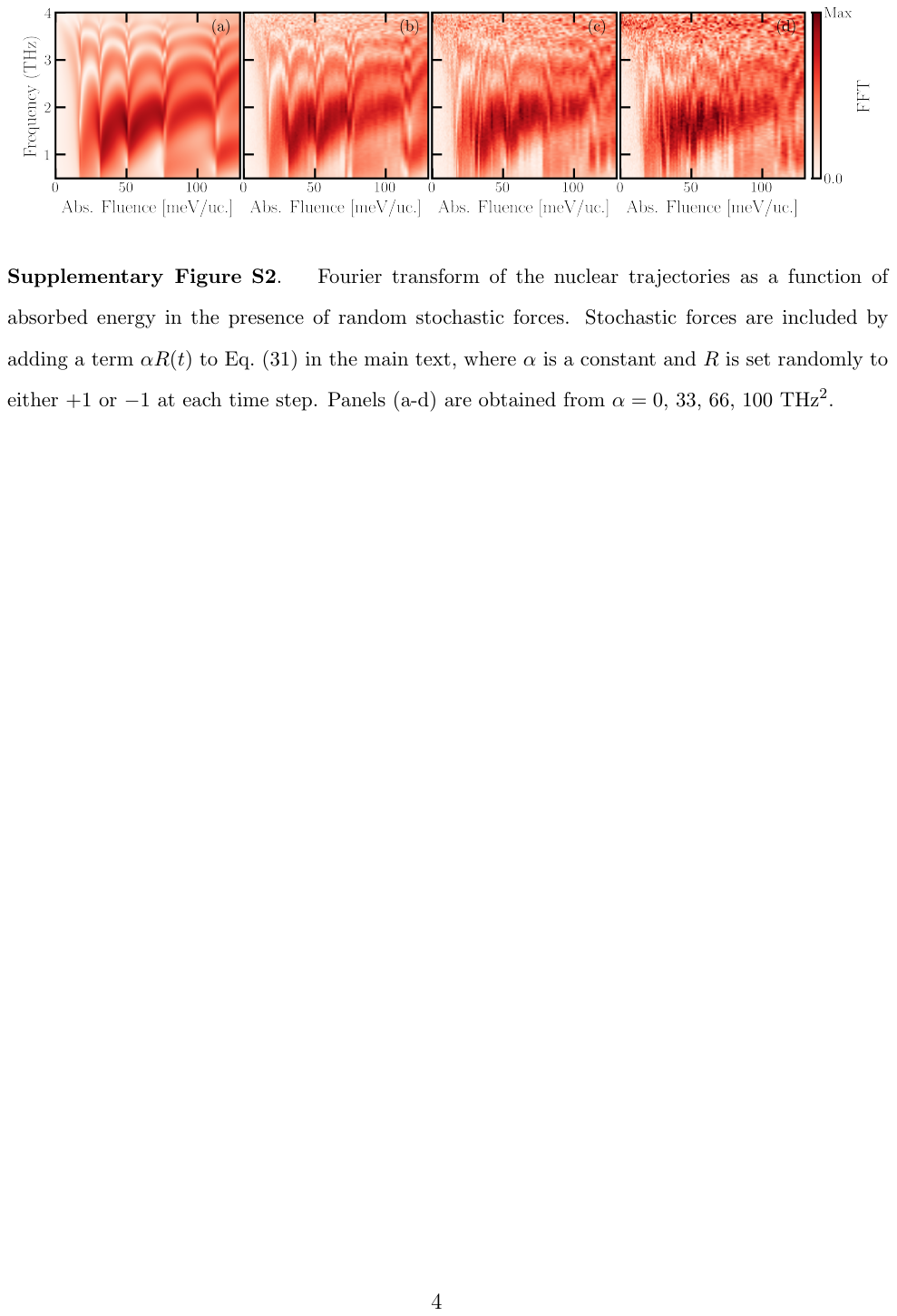}
\end{figure}

\begin{figure}[htpb]
    \centering
    \includegraphics[width=0.99\textwidth]{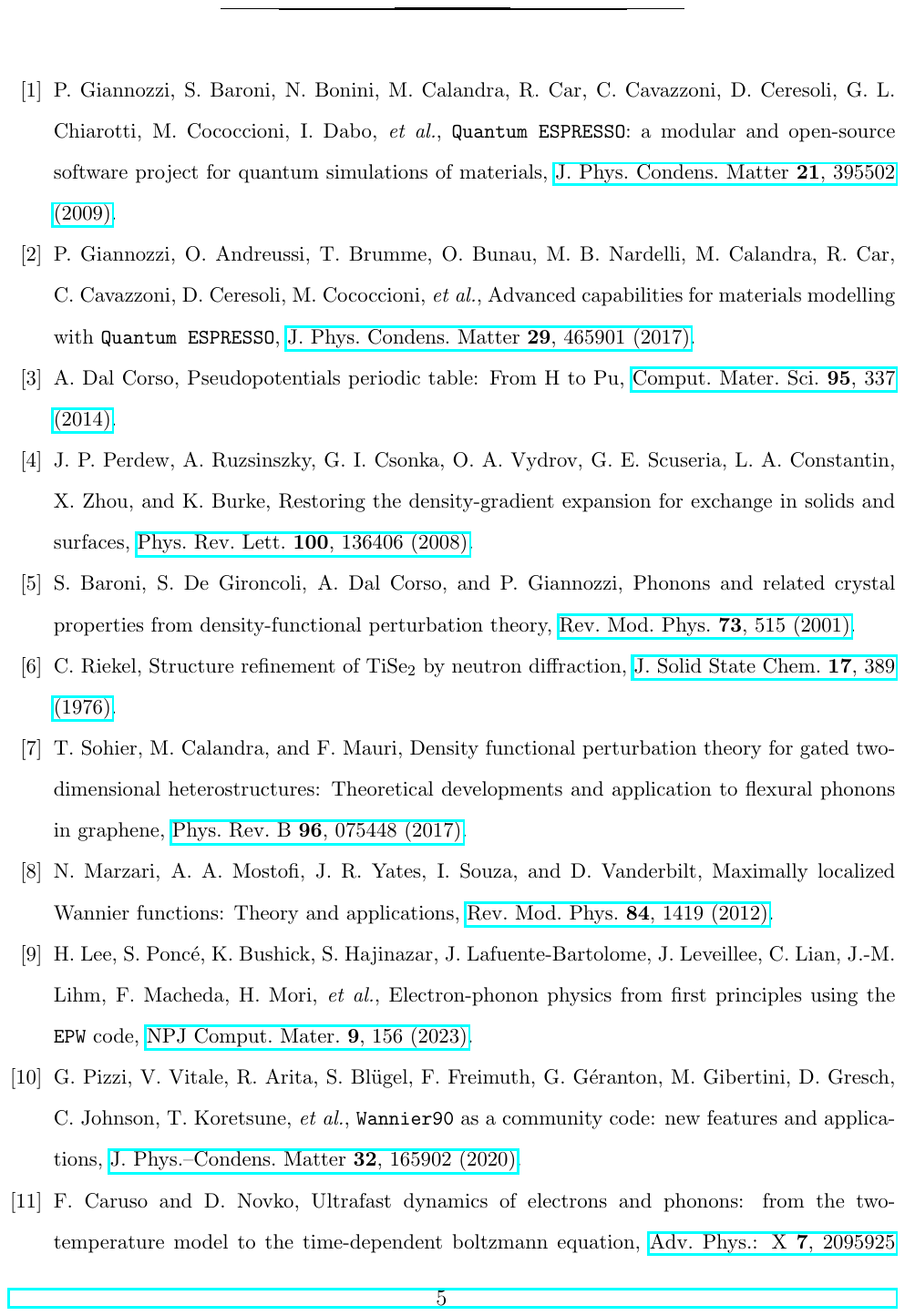}
\end{figure}

\begin{figure}[htpb]
    \centering
    \includegraphics[width=0.99\textwidth]{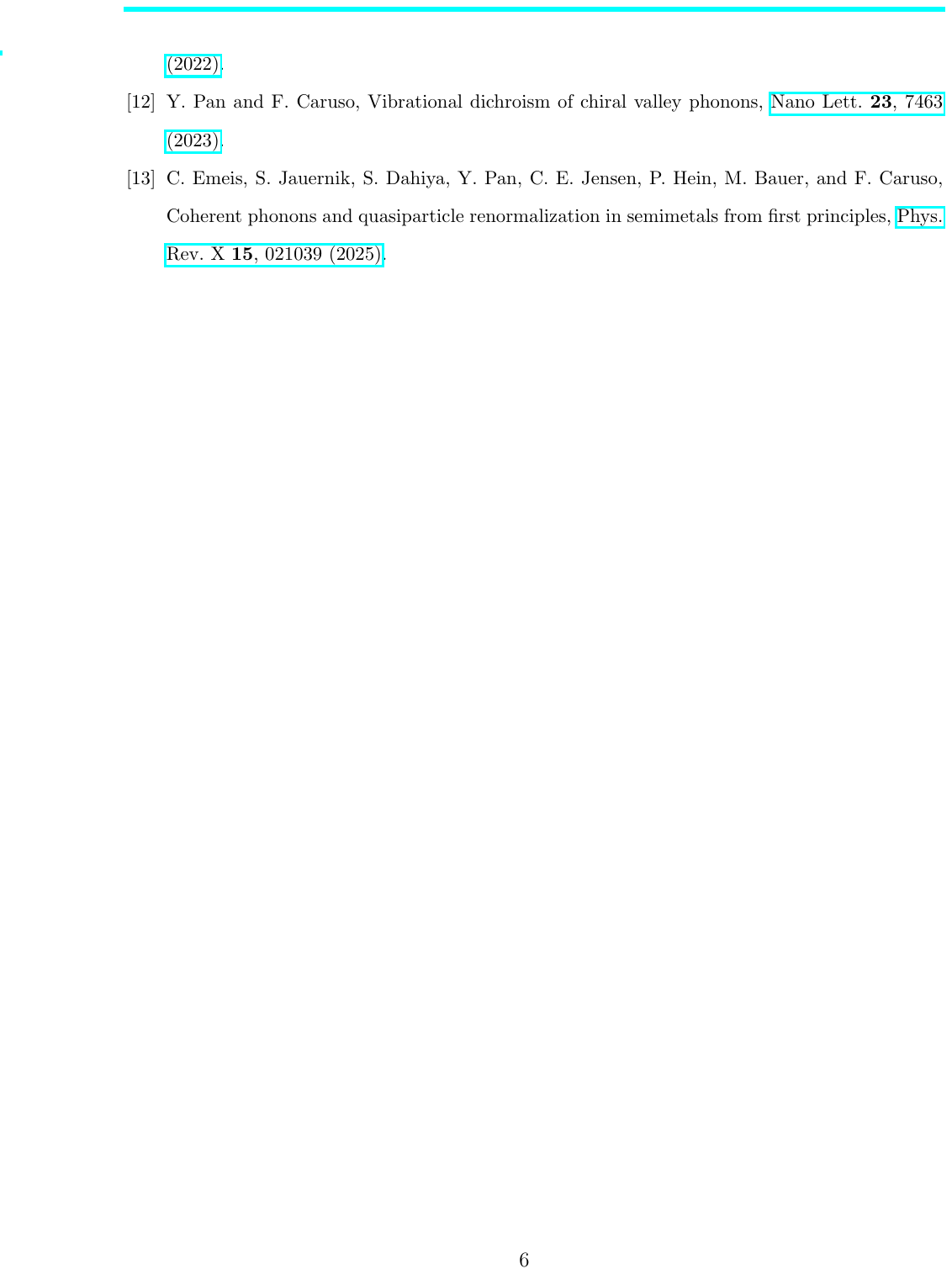}
\end{figure}

\end{document}